\documentclass[twocolumn,english,showpacs,preprintnumbers,
amsmath,amssymb,floatfix,nofootinbib]{revtex4-1}

\usepackage[T1]{fontenc}
\usepackage[latin9]{inputenc}
\usepackage{color}
\usepackage{array}
\usepackage{amstext}
\usepackage{graphicx}
\usepackage{esint}
\usepackage{rotating}
\usepackage{slashed}
\usepackage{siunitx}
\usepackage[font=small,labelfont=bf]{caption}
\usepackage{soul,xcolor}
\usepackage{xcolor}
\usepackage[colorlinks = true,
linkcolor = magenta,
urlcolor  = blue,
citecolor = red,
anchorcolor = blue]{hyperref}
\usepackage{ulem}
\usepackage[titles]{tocloft}

\makeatletter

\begin{document}

\setstcolor{blue}


%
%
\title{ 
Fragmentation Functions for $\Xi ^-/\bar{\Xi}^+$ 
Using Neural Networks 
} 
%
%

\author{Maryam Soleymaninia$^{1}$}
\email{Maryam\_Soleymaninia@ipm.ir}

\author{Hadi Hashamipour$^{1}$}
\email{H\_Hashamipour@ipm.ir}

\author{Hamzeh Khanpour$^{2,3,4,1}$}
\email{Hamzeh.Khanpour@cern.ch}

\author{Hubert Spiesberger$^5$}
\email{spiesber@uni-mainz.de}

\affiliation {
$^{1}$School of Particles and Accelerators, 
Institute for Research in Fundamental Sciences (IPM), 
P.O.Box 19395-5531, Tehran, Iran \\
$^{2}$Dipartimento Politecnico di Ingegneria ed Architettura, 
University of Udine, Via della Scienze 206, 33100 Udine, Italy \\
$^{3}$International Centre for Theoretical Physics (ICTP), Strada Costiera 11, 34151 Trieste, Italy \\
$^{4}$Department of Physics, University of Science and Technology 
of Mazandaran, P.O.Box 48518-78195, Behshahr, Iran  \\
$^{5}$PRISMA$^{+}$ Cluster of Excellence, 
Institut f\"ur Physik, Johannes-Gutenberg-Universit\"at, 
Staudinger Weg 7, D-55099 Mainz, Germany
}

\date{\today}

%

\begin{abstract}

We present a determination of fragmentation functions (FFs) for 
the octet baryon $\Xi ^-/\bar{\Xi}^+$ from data for single 
inclusive electron-positron annihilation. Our parametrization 
in this QCD analysis is provided in terms of a Neural Network (NN). 
We determine fragmentation functions for $\Xi ^-/\bar{\Xi}^+$ 
at next-to-leading order and for the first time at 
next-to-next-to-leading order in perturbative QCD. We discuss 
the improvement of higher-order QCD corrections, the quality of 
fit, and the comparison of our theoretical results with the 
fitted datasets. As an application of our new set of 
fragmentation functions, named {\tt SHKS22}, we present 
predictions for $\Xi ^- / \bar{\Xi}^+$ baryon production 
in proton-proton collisions at the LHC experiments. 

\end{abstract}
%


\maketitle
\tableofcontents{}


%
\section{Introduction}
\label{sec:introduction}
%

Fragmentation Functions (FFs) are necessary to describe the 
hadronization process in which partons turn into hadrons in 
the final state. They are needed to describe hadron production  
in semi-inclusive processes in electron-positron annihilation 
(SIA), and, together with parton distribution functions (PDFs) 
in lepton nucleon scattering and (anti)proton-proton collisions. 
Parametrizations of FFs for light hadrons, as well as for mesons 
with charm and bottom quarks, are available since long and the 
corresponding phenomenology is well established. New improved 
data are routinely used to improve the existing FF parametrizations, 
see for example~\cite{Khalek:2021gxf,Abdolmaleki:2021yjf} and 
references therein. In recent years, progress has been made also 
for more specific applications, as for example the determination 
of transverse momentum dependent 
FFs~\cite{Kang:2015msa,Soleymaninia:2019jqo}, FFs for heavier 
hadrons, like $\Lambda_c^\pm$ baryons~\cite{Kniehl:2020szu}
and medium modified FFs~\cite{Zurita:2021kli}. With more 
precise data it seems also possible to aim at a combined 
determination of PDFs and FFs~\cite{Moffat:2021dji} and future 
experiments, for example at the EIC~\cite{Aschenauer:2019kzf}, 
will eventually improve our understanding of FFs.

In this work we are interested in a determination of FFs 
for the production of $\Xi^-$ and its anti-particle, i.e.\ 
the double-strangeness carrying octet baryon. On the 
experimental side, most important information about 
$\Xi^-$ production is coming from SIA 
measurements in $e^+ e^-$ annihilation on the $Z$-boson 
resonance at the LEP experiments ALPEH~\cite{ALEPH:1996oqp}, 
DELPHI~\cite{DELPHI:1995dso} and OPAL~\cite{OPAL:1996gsw}. 
Earlier data are also available from MARKII~\cite{Klein:1986ws} 
and TASSO~\cite{TASSO:1983qye,TASSO:1988qlu} at somewhat 
smaller center-of-mass energies. In recent years, also data 
for $\Xi^-$-production in $pp$ and $\bar{p}p$ collisions 
have been published, covering center-of-mass energies from 
$\sqrt{s}=17.3$~GeV at NA61/SHINE \cite{NA61SHINE:2020dwg}, 
$\sqrt{s}=200$~GeV at STAR~\cite{STAR:2006nmo,Heinz:2007ci}, 
1.96~TeV at CDF~\cite{CDF:2011dvx}, and up to 7 and 13~TeV 
at CMS~\cite{CMS:2011jlm,CMS:2019isl} and 
ALICE~\cite{ALICE:2012yqk,ALICE:2019avo,ALICE:2020jsh}. 
The relevant energy scale in these collider experiments is the 
transverse momentum of the produced hadron which is, however, 
rather low and does not reach values above 10~GeV. 

On the theoretical side, the only available analysis of the 
FF of the $\Xi^-$ was made in a statistical approach and used 
data from $e^+e^-$ annihilation~\cite{Bourrely:2003wi}. This 
motivates us to perform a new determination of the $\Xi^-$ FF 
using the by now well established technique of neural networks. 

The main motivation of our present study is to determine a new 
set of $\Xi^- /\bar{\Xi}^+$ FFs. 
The universality property of 
FFs allows us to perform a QCD analysis using all available 
measurements to determine the FFs for a specific hadron.
The FFs extracted in this way can be used to obtain theoretical 
predictions for other processes, like hadron production in 
{\it pp} collisions. In this paper we present a determination 
of the FFs of $\Xi^- /\bar{\Xi}^+$ baryon in which single 
inclusive electron-positron annihilation (SIA) data are analyzed 
at next-to-leading order (NLO) and next-to-next-to-leading 
order (NNLO) accuracy in perturbative QCD. 

We hope that our determination of $\Xi^- /\bar{\Xi}^+$ FFs 
will provide the necessary ingredient needed to obtain 
predictions for future measurements in a well-defined 
reference framework based on independent parton-to-hadron 
fragmentation in perturbative QCD. Such a reference will 
be needed to test theoretical ideas about mechanisms for 
hadron production. With future precise data, a comparison 
with predictions from perturbative QCD is expected to 
reveal non-perturbative mechanisms and possible collective 
phenomena in a high-temperature quark-gluon plasma (QGP). 
For instance, one of the signatures for creating a QGP is 
production of strange hadrons in heavy-ion collisions
\cite{Rafelski:1982pu,Koch:2017pda}. In particular hadrons 
with low transverse momentum are expected to reflect the 
properties of the bulk system such as collective expansion, 
the hadro-chemical composition, and the temperature at 
freeze-out. Measurements of identified hadrons at low 
transverse momentum used to extract these properties 
\cite{ALICE:2013mez,NA44:1996xlh} will also profit from 
theory predictions for higher transverse momenta obtained 
in a reference framework.

We use the recent publicly available {\tt MontBlanc} 
package~\cite{MontBlanc} in our analysis. This package is based 
on neural networks which are believed to provide a parametrization 
with minimal bias. This allows us to control the precision of the 
$\Xi^- /\bar{\Xi}^+$ FFs, based on an efficient minimization 
algorithm. The numerical calculations of the hadron production 
cross-section in $e^+e^-$ annihilation and the scale evolution 
of the FFs is also performed in this framework. The fitting 
methodology in the framework of this package has been developed 
by the NNPDF 
Collaboration~\cite{NNPDF:2017mvq,Khalek:2021gxf,Bertone:2017tyb}.

We will discuss in detail the novel aspects of the methodology 
used in {\tt SHKS22} analysis, the fit quality, the perturbative 
convergence upon inclusion of higher-order QCD correction, 
and the stability upon variations of the kinematic cuts applied 
to the SIA dataset. We will also show that the inclusion of 
higher-order QCD corrections in our analysis improves the 
description of the analyzed SIA data.

The structure of the paper is as follows: In Sec.~\ref{sec:QCD} 
we review the theoretical formalism for inclusive hadron 
production in electron-positron annihilation. In Sec.~\ref{sec:methodology} our methodology based on a neural 
network (NN) framework is presented. We illustrate the Monte 
Carlo methodology adopted in our analysis to calculate the 
uncertainties of FFs and the optimal fit in 
Sec.~\ref{sec:MC}. In Section~\ref{sec:data} 
we describe our selection of SIA experimental data included in 
this study. Details of our new {\tt SHKS22} FFs are 
presented in Sec.~\ref{sec:results}. We also discuss in this 
section the impact of higher-order perturbative QCD (pQCD) 
corrections and compare numerical results for the differential 
cross sections with the analyzed experimental datasets. In 
Sec.~\ref{sec:LHC-RHIC}, we present and discuss some predictions 
for possible future measurements at hadron colliders. Finally, 
we summarize our conclusions in Sec.~\ref{sec:conclusion}.

%
\section{QCD Framework}
\label{sec:QCD}
%

For single inclusive hadron production in electron-positron 
annihilation, the differential cross-section is given by the 
formula 
\begin{equation}
\label{diffcross}
\frac{d \sigma^{\it h}} {dz}(z, Q^2) =
\frac{4\pi \alpha ^2 (Q)} {Q^2}
 F^{\it h}(z, Q^2) \,,
\end{equation}
where $F^h$ is the fragmentation structure function, and 
$\alpha$ is the QED running coupling\footnote{ 
   We follow the convention of 
   Ref.~\cite{ParticleDataGroup:2018ovx} where the combination 
   of the transverse and longitudinal fragmentation structure 
   functions $F^{\it h} = F_T^{\it h} + F_L^{\it h}$ is used. 
   This is the combination which can be determined after 
   inegrating over the hadron's emission angle with respect 
   to the beam direction. Sometimes $F^{\it h}$ is also 
   denoted $F_2^{\it h}$ in the literature, see e.g.\ 
   Ref.~\cite{Bertone:2017tyb}. 
}. 
We follow the standard collinear factorization framework 
\cite{esw-book} where the QCD cross-sections can be factorized 
into perturbatively calculable partonic hard cross sections
and non-perturbative distribution functions. Thus the structure 
function is given as a convolution of FFs and hard-scattering 
coefficient functions:
\begin{eqnarray}
\label{eq:structure-functions}
F^{\it h}(z, Q^2) 
&=& 
\frac{1}{n_f} \sum_q^{n_f}\hat{e}^2_q(Q) \times 
\\
&& 
\left[D^{\it h}_S(z, Q^2)\otimes C^S_{2,q}(z, \alpha _s(Q)) \right. 
\nonumber \\
&& 
+ D^h_g(z, Q^2) \otimes C^S_{2, g}(z, \alpha _s(Q)) 
\nonumber \\
&& 
\left.
+ D^{\it h}_{\rm NS}(z, Q^2) \otimes C^{\rm NS}_{2, q}(z, \alpha _s(Q)) 
\right] 
\, , 
\nonumber
\end{eqnarray}
where $\hat{e}_q(Q)$ are scale-dependent quark electroweak charge 
factors. Their definition can be found in 
Ref.~\cite{Rijken:1996ns}. In 
Eq.~(\ref{eq:structure-functions}), the function $D^h_g(z, Q^2)$ 
represents the gluon FF, and $D^{\it h}_S(z, Q^2)$ and 
$D^{\it h}_{\rm NS}(z, Q^2)$ are the singlet and non-singlet 
combinations of FFs which can be written as 
\begin{eqnarray}
\label{eq:singlet}
D_S^{\it h}(z, Q^2) 
&=& 
\sum_q^{n_f} D^{\it h}_{q^+}(z, Q^2) \, , 
\\
\label{eq:nonsinglet}
D_{\rm NS}^{\it h}(z, Q^2) 
&=&
\sum_q^{n_f} \frac{\hat{e}^2_q}{\left<\hat{e}^2\right>} 
\left[D^{\it h}_{q^+}(z, Q^2) - D_S^{\it h}(z, Q^2)\right],
\end{eqnarray}
with $D^h_{q^+} = D^h_{q} + D^h_{\bar{q}}$. 
The coefficient functions $C^{S,NS}_{q,g}(z, \alpha_s(Q))$ 
for the singlet and non-singlet combinations have been computed 
up to $\cal{O}$ $(\alpha_s^2)$~\cite{Blumlein:2006rr,Rijken:1996npa}.
The average of the effective quark electroweak charges
over the number of active flavors reads 
\begin{eqnarray}
\left<\hat{e}^2\right>= 
\frac{1}{n_f}\sum_q^{n_f}\hat{e}^2_q(Q).
\end{eqnarray}

The evolution of FFs with the energy scale $Q$ is performed 
by the standard DGLAP evolution 
equation~\cite{Gribov:1972ri,Altarelli:1977zs,Dokshitzer:1977sg}.
The evolution of the singlet combination of FFs mixes with the 
gluon FF and is given by 
\begin{eqnarray}
  \frac{\partial}{\partial\ln Q^2}
  \left(
    \begin{array}{c}
      D_{S}^h 
      \\
      D_g^h
    \end{array}
  \right)(z,Q^2)
&=&
\left(
  \begin{array}{cc}
    P^{qq} & 2n_fP^{gq} 
    \\
    \frac{1}{2n_f}P^{qg} & P^{gg}
  \end{array}
\right)\left(z,\alpha_s\right) 
\nonumber\\ 
&\otimes& 
\left(
\begin{array}{c}
D_S^h 
\\
D_g^h
\end{array}
\right)(z,Q^2)
\,\mbox{,}
\label{eq:evsingletg}
\end{eqnarray}
while for the nonsinglet combination of FFs, the DGLAP equation 
can be written as
\begin{eqnarray}
  \frac{\partial}{\partial\ln Q^2} D_{\rm NS}^h(z,Q^2)
  =
  P^+\left(z,\alpha_s\right)\otimes D_{\rm NS}^h(z,Q^2)
\label{eq:evNS}
\,\mbox{,}
\end{eqnarray}
where $P^{ji}$ and $P^+$ are the time-like splitting 
functions which have a perturbative expansion in 
terms of the strong coupling constant $\alpha_s$,
\begin{eqnarray}
  P^{ji,+}\left(z,\alpha_s\right)
  =
  \sum_{l=0}a_s^{l+1}\, P^{ji,+\,(l)}(z)
  \,\mbox{,}
\label{eq:pert}
\end{eqnarray}
where $i,j=g,q$ and $a_s \equiv \alpha_s/(4\pi)$. The time-like 
splitting functions have been calculated up to $\mathcal{O}(a_s^3)$ 
in the $\overline{\rm MS}$ scheme and can be found 
in Refs.~\cite{Mitov:2006ic,Moch:2007tx,Almasy:2011eq}.

The calculation of FFs in our analysis is defined in the 
zero-mass variable-flavor-number scheme (ZM-VFNS) where 
all active flavors are treated as massless quarks. Nevertheless, 
 a dependence on the heavy-quark masses enters through 
the fact that the FFs for heavy-quark contributions exhibit 
a threshold: at scales below the heavy-quark mass the scale 
evolution is stopped and the FFs are kept constant. We use 
$m_c = 1.51$~GeV and $m_b = 4.92$~GeV for the charm and bottom 
thresholds, respectively. We choose  $\alpha_s(M_Z) = 0.118$ as 
a reference value which is close to the world-average of the 
Particle Data Group~\cite{ParticleDataGroup:2018ovx}.

%
\section{Methodology and Input Parameterization}
\label{sec:methodology}
%

In this section we present the methodology applied in the 
{\tt SHKS22} analysis, namely the neural network (NN) 
technique, and describe the basic assumption for the input 
FFs parameterization. A neural network can be viewed as a 
complicated prescription to parametrize a function. The 
architecture of the network, i.e.\ in particular the number 
of layers and nodes and the way how nodes are connected, 
determines the complexity of this parametrization. The 
typical number of parameters needed to fix a NN is around 
100, or more. This is much larger than the number of 
parameters used in conventional approaches, where simple 
functional forms for FFs requires to specify in the order 
of 15 to 30 parameters. Therefore, NNs are expected to be 
much more flexible and the bias due to the choice of a 
specific functional form is greatly reduced. For more 
detailed information about the neural network approach 
we refer to Refs.~\cite{Bertone:2017tyb,Khalek:2021gxf}.

The $\Xi^- + \bar{\Xi}^+$ fragmentation functions for a 
parton flavor of flavor combination $i$ in terms 
of a neural network is defined at the initial scale 
$Q_0 = 5$ GeV by 
\begin{eqnarray}
\label{NN}
zD^{\Xi^- + \bar{\Xi}^+}_i(z,Q_0)
= 
(N_i(z,\theta) - N_i(1,\theta))^2 \, ,
\end{eqnarray}
where $N_i(z,\theta)$ is a one-layered feed-forward neural 
network, and $\theta$ denotes the parameter set. The neural 
network architecture is $[I,N,O]$ where $I$ is the number of 
input nodes, $N$ denotes the number of intermediate nodes and 
$O$ is the number of output nodes. We choose a neural network 
with a single input node describing the scaling variable $z$, 
a sigmoid activation function for $20$ intermediate nodes and 
a linear activation function for $3$, $5$ or $6$ output nodes 
corresponding to the flavor combinations in three different 
FF sets, I, II or III, respectively, which we will introduce 
in the following.

SIA experimental measurements are available only for the 
production of the sum of $\Xi^-$ and its antiparticle 
$\bar{\Xi}^+$. The fragmentation structure functions in 
Eq.~(\ref{eq:structure-functions}) do not distinguish between 
quark and antiquark FFs, i.e.\ we can only determine the 
combinations $D^h_{q^+} = D^h_{q} + D^h_{\bar{q}}$ with 
$h = \Xi^- + \bar{\Xi}^+$. The singlet and non-singlet 
combinations and the gluon FF enter with different coefficients 
in the SIA cross section and a separation of these three 
components can be expected provided data for the $z$ and $Q$ 
dependence are precise enough. 

Our baseline FF parametrization, set I, is chosen assuming 
the following flavor combinations: 
\begin{eqnarray}
\label{SETI}
zD^{\Xi^- + \bar{\Xi}^+}_{d^+ + s^+},~ 
zD^{\Xi^- + \bar{\Xi}^+}_{u^+, c^+, b^+},~ 
zD^{\Xi^- + \bar{\Xi}^+}_g 
\, .
\end{eqnarray}
Since the favored quark flavors in the hadron state $\Xi ^-$ 
are $(dss)$ and the value of the corresponding electroweak 
charges $\hat{e}_q$ for $d$ and $s$ quarks are the same, 
we adopt the specific combination $d^+ + s^+$, as was done 
also in Ref.~\cite{Binnewies:1994ju}. In addition we assume 
complete symmetry between the disfavored quarks $u^+$, $c^+$ 
and $b^+$. Therefore, in total we choose $3$ independent 
flavor combinations corresponding to the neural network 
architecture $[1, 20, 3]$. 

The FFs for up- and down-type quarks can be separated 
because they contribute with different charge factors 
to the cross section. A separation of the heavy-quark 
components will be possible if charm- or bottom-tagged 
data become available. 
At present, there are no heavy-flavor tagged data and 
it is unlikely that the precision is high enough to be 
useful for a further separation of flavors, beyond our 
ansatz described above. Nevertheless, as a test of our 
approach, we will also study two extended sets, namely 
set II, assuming the following combinations of FFs:
\begin{eqnarray}
\label{SETII}
zD^{\Xi^- + \bar{\Xi}^+}_{d^+ + s^+},~ 
zD^{\Xi^- + \bar{\Xi}^+}_{u^+},~ 
zD^{\Xi^- + \bar{\Xi}^+}_{c^+},~ 
zD^{\Xi^- + \bar{\Xi}^+}_{b^+},~ 
zD^{\Xi^- + \bar{\Xi}^+}_g, 
\nonumber
\\
\end{eqnarray}
where, in addition to the light quarks and the gluon, we assume 
separate parametrizations for the heavy quarks $c^+$ and $b^+$, 
and set III where all flavors are described by different FFs: 
\begin{eqnarray}
\label{SETIII}
&&
zD^{\Xi ^-/\bar{\Xi}^+}_{d^+}, 
zD^{\Xi ^-/\bar{\Xi}^+}_{u^+}, 
zD^{\Xi ^-/\bar{\Xi}^+}_{s^+}, 
zD^{\Xi ^-/\bar{\Xi}^+}_{c^+}, 
zD^{\Xi ^-/\bar{\Xi}^+}_{b^+}, 
\nonumber\\
&&
zD^{\Xi ^-/\bar{\Xi}^+}_{g}
\end{eqnarray}
Thus, for set II (III) we increase the number of independent 
distribution functions to $5$ ($6$) and the NN architecture 
is $[1, 20, 5]$ ($[1,20,6]$). We study sets II and III in order 
to investigate how well FFs are restricted by data and which 
uncertainty could be generated by choosing a specific, 
possibly too general, parametrization. One can find similar 
approaches in the previous literature. 

The number of parameters needed to specify the chosen 
NN architectures are $103$, $145$ and $166$ for sets 
I, II and III, respectively. Due to the redundancy of 
these parameters inherent to the NN architecture by 
construction, the number of independent parameters is, 
however, much smaller and usually difficult to determine. 
Therefore we will not quote $\chi^2$ values per degree of 
freedom, but rather follow the general convention and show 
$\chi^2$ normalized to the number of data points when we 
compare the fit quality in the following.

%
\section{Monte Carlo Procedure and Determination of the Optimal Fit}
\label{sec:MC}
%

There are mainly two methods used in the literature to obtain 
a reliable determination of the uncertainties of FFs. One of 
them is the iterative Hessian approach which has been used 
for example in the {\tt DEHSS} 
analysis~\cite{deFlorian:2014xna,deFlorian:2017lwf} and has been 
developed in Refs.~\cite{Pumplin:2000vx,Pumplin:2001ct}. 
The other important one, developed in Ref.~\cite{Sato:2016tuz}, 
is the Monte Carlo procedure used in the recent 
{\tt JAM}~\cite{Moffat:2021dji,Sato:2016wqj} and 
{\tt NNFF}~\cite{Bertone:2017tyb} fits. We adopt the statistical 
framework in our {\tt SHKS22} analysis. 
 
The {\tt MontBlanc} framework~\cite{Khalek:2021gxf} utilizes 
Monte Carlo together with NNs to extract information about the 
FFs and their uncertainty from the experimental data. 
The available data are used to generate a discrete sample of 
FF values, and NNs are employed to interpolate between data  
points which is needed in order to approximate the required 
integrations. The first step is training of the NN and creating 
$N_\mathrm{rep}$ replicas of NNs corresponding to pseudo-datasets. 
Minimization of an appropriately chosen $\chi^2$ allows one to 
choose NNs such that they reproduce a probability distribution 
of measured points and theoretical predictions with the same 
mean, variance and correlation as the experimental data. We 
refer the reader to 
Refs.~\cite{Forte:2002fg,Sato:2016wqj,Khalek:2021gxf,Bertone:2017tyb} 
and references therein for a detailed discussion of the approach. 

The Monte Carlo approach assumes that the experimental data 
follow a multivariate Gaussian distribution, 
\begin{align}
\mathcal{G}(\boldsymbol{x}^k) \propto \exp 
\left( \left(\boldsymbol{x}^k - 
\boldsymbol{\mu}\right)^T 
\cdot \boldsymbol{C}^{-1} \left(\boldsymbol{x}^k - 
\boldsymbol{\mu}\right)  \right),
\end{align}
where $\boldsymbol{x}^k = \left\{x_1^k , x_2^k, \cdots, 
x_{N_{\mathrm{dat}}}^k \right\}$ are so called ``replicas'', 
i.e.\ pseudo-datasets for a set of $N_{\mathrm{dat}}$ measured 
data points. The vector of expectation values of this distribution 
is $\boldsymbol{\mu}$ which corresponds to experimental data, 
and $\boldsymbol{C}$ is the covariance matrix of the data which 
contains all information on the uncertainties and correlations. 
The elements of the covariance matrix are defined as in 
Ref.~\cite{Khalek:2021gxf},
\begin{align}
C_{ij} = \delta_{ij} 
\sigma_{i,\mathrm{unc}}^2 + \sum_{\beta} 
\sigma_{i,\mathrm{corr}}^{(\beta)}  
\sigma_{j,\mathrm{corr}}^{(\beta)},
\end{align}
where $\sigma_{i,\mathrm{unc}}^2$ denotes the sum of squares 
of all uncorrelated uncertainties, such as the statistical 
and systematic uncertainties for the $i$th data point and 
$\sigma_{i,\mathrm{corr}}^{(\beta)}$ is the correlated 
uncertainty due to $\beta$ source for the same point.

The Monte Carlo approach for the uncertainty propagation uses 
$N_\mathrm{rep}$ replicas of the measured data, $\boldsymbol{x}^k$, 
to turn the experimental uncertainty into that of the NNs that 
parametrize the FFs. For this purpose, the covariance matrix 
$\boldsymbol{C}$ is decomposed to a lower triangular 
$\boldsymbol{L}$ using Cholesky method ($\boldsymbol{C} = 
\boldsymbol{L} \cdot \boldsymbol{L}^T$). If one applies this 
matrix to an $N_\mathrm{dat}$-dimensional Gaussian random 
vector, $\boldsymbol{r}^k$, a vector with the covariance 
properties of the experimental data is obtained. Accordingly 
the pseudo dataset is constructed as follows (by 
\texttt{Ceres Solver}~\cite{ceres-solver}),
\begin{equation}
\boldsymbol{x}^k = 
\boldsymbol{\mu} + 
\boldsymbol{L} 
\cdot \boldsymbol{r}^k
\end{equation}
This procedure ensures that for sufficiently large 
$N_\mathrm{rep}$, the set of replicas satisfies the following 
conditions~\cite{Khalek:2021gxf},
\begin{equation}
\frac{1}{N_\mathrm{rep}}
\sum_{k}^{N_\mathrm{rep}} 
{x}^k_i \simeq \mu_i 
\, , \quad  
\frac{1}{N_\mathrm{rep}}
\sum_{k}^{N_\mathrm{rep}} 
{x}^k_i {x}^k_j \simeq \mu_i \mu_j + C_{ij} \, .
\end{equation}
At the end of the QCD analysis, a set of $N_\mathrm{rep}$ 
trained NNs are gained and this allows us to estimate the 
functional integral by averaging over the set of replicas. 
Notably, estimates for expectation value, uncertainty and 
correlation of the distribution of an observable $\mathcal{D}$, 
e.g.\ a cross section, is given by~\cite{Forte:2002fg},
\begin{align}
&
\left<\mathcal{D}(x)\right> 
= 
\frac{1}{N_\mathrm{rep}} \sum_{k=1}^{N_\mathrm{rep}} \mathcal{D}^k(x) 
\, , 
\nonumber \\
&
\sigma_\mathcal{D} 
= 
\sqrt{ \left< \mathcal{D}^2  \right> 
      - \left< \mathcal{D}  \right>^2 } \, , 
\nonumber \\
&c_{\mathcal{D}}^{ij} 
= 
\frac{ \left< \mathcal{D}_i \mathcal{D}_j \right> 
      - \left< \mathcal{D}_i \right> \left< \mathcal{D}_j \right>}
{\sigma_i \sigma_j}
\, . 
\end{align}
In the {\tt SHKS22} FFs analysis, we have chosen to include 
$N_\mathrm{rep} = 300$, although already a smaller number of 
replicas, say 200, could be enough to obtain a sub $1\%$ 
accuracy, see Ref.~\cite{Khalek:2021gxf}.

%
\section{Experimental Data}
\label{sec:data}
%

\begin{table*}[tb]
\renewcommand{\arraystretch}{2}
\centering 
\scriptsize
\begin{tabular}{lrccccc}	
\\
\hline
Experiment 
   & ~ $\sqrt{s}$ ~ 
   & ~ $N_{\mathrm{dat}}(z>z_{min})$ ~
   & ~ $\chi^2/N_{\mathrm{dat}}$ Set I(NLO) 
   & ~ $\chi^2/N_{\mathrm{dat}}$ Set II(NLO) 
   & ~ $\chi^2/N_{\mathrm{dat}}$ Set III(NLO) 
   & ~ $\chi^2/N_{\mathrm{dat}}$ Set I(NNLO) 
\rule[-3mm]{0mm}{8mm}
\\
\hline \hline
	{\tt ALEPH}~\cite{ALEPH:1996oqp}   
	& 91.2 & 8 & 1.860 & 2.039 & 2.035 & 1.714 
	\\
	{\tt DELPHI}~\cite{DELPHI:1995dso} 
	& 91.2 & 4 & 0.659 & 0.660 & 0.668 & 0.673 
	\\
	{\tt OPAL}~\cite{OPAL:1996gsw} 
	& 91.2 & 5 & 1.552 & 1.682 & 1.726 & 1.434 
	\\
	{\tt MARKII}~\cite{Klein:1986ws} 
	& 29   & 4 & 2.348 & 2.399 & 2.428 & 2.171 
	\\
	{\tt TASSO-34}~\cite{TASSO:1983qye} 
	& 34   & 3 & 1.747 & 1.805 & 1.807 & 1.700 
	\\
	{\tt TASSO-34.8}~\cite{TASSO:1988qlu} 
	& 34.8 & 3 & 0.810 & 0.731 & 0.777 & 0.891 
	\\
\hline \hline
Total $ \chi^2/N_{\mathrm{dat}}$ & & 27 & 1.540 & 1.631 & 1.643 & 1.451  
\\
\hline \hline 
\\
\end{tabular}
\caption{\small 
The list of input datasets for $\Xi ^- + \bar{\Xi}^+$ 
production included in the {\tt SHKS22} analysis. For each 
dataset we indicate the corresponding reference and the 
center-of-mass energy $\sqrt{s}$. The number of data points 
$N_{\mathrm{dat}}$ that satisfy the kinematic cut ($z>0.1$) 
is displayed in the third column. In columns 4, 5, and 6 we 
show the value of $\chi^2$ per point resulting from the FF 
fit at NLO for set I, set II and set III, respectively. 
The last column shows, correspondingly, the $\chi^2$ values 
for the NNLO fit. The last row contains the total values 
of $\chi^2$ divided by the number of data points. 
}
\label{tab:datasets}
\end{table*}

The data included in our analysis is obtained from SIA 
measurements of the sum of cross sections for $\Xi^-$ and 
$\bar{\Xi}^+$ production. We make use of all available 
experimental data reported by the ALEPH~\cite{ALEPH:1996oqp}, 
DELPHI~\cite{DELPHI:1995dso} and 
OPAL~\cite{OPAL:1996gsw} Collaborations at CERN, the MARKII 
Collaboration~\cite{Klein:1986ws} at SLAC, and the TASSO 
Collaboration~\cite{TASSO:1983qye,TASSO:1988qlu} at DESY. 
The datasets are summarized in Table~\ref{tab:datasets}. 
All the data are for inclusive untagged cross-section 
measurements. The ALEPH, DELPHI and OPAL data are presented 
as multiplicities $1/{\sigma_{tot}}d\sigma /dx_p$ normalized 
to the total cross-section at the center-of-mass energy of 
$\sqrt{s}=M_Z$. Here, $x_p = 2|\bold{P_h}|/\sqrt{s}$ is
the scaled hadron three-momentum. Alternatively, one can 
also use $z = 2E_h/\sqrt{s}$, i.e.\ the energy of the hadron 
$h$ scaled to the beam energy $\sqrt{s}/2$. The relation between 
these two scaling variables is given by 
\begin{equation}
\label{variables}
z = x_p \sqrt{1 + \frac{4}{x_p^2}\frac{m_h^2}{s}} 
= 2 \sqrt{\frac{m_h^2 + |\bold{P_h}|^2}{s}}.
\end{equation}
The data from the MARKII and TASSO Collaborations are given in 
different formats. While the TASSO-34.8 data are differential 
cross-sections $d\sigma/dp_h$ measured at the center-of-mass 
energy $\sqrt{s} = 34.8$~GeV, the observables for the MARKII 
and TASSO-34 data are rescaled cross sections and reported 
for $(s / \beta) d\sigma/dz$ at the center-of-mass energies 
$\sqrt{s} = 29$~GeV and $\sqrt{s}=34$~GeV, respectively. The 
parameter $\beta = |\bold{P_h}|/E_h$ is the velocity of the 
final state hadron $h$.

\begin{table*}[t!]
\renewcommand{\arraystretch}{2}
\centering 	
\scriptsize
\begin{tabular}{lccccr}	
\\
\hline
$z_{min}$  cut ~&~  number of data points~ & ~$\chi^2/N_{\mathrm{dat}}$ 
\\
\hline \hline
0.02   & 36  & 4.412 \\ 
0.05   & 34  & 1.967 \\ 
0.075  & 31  & 2.010 \\ 
0.1    & 27  & 1.540 \\ 
\hline \hline 
\\
\end{tabular}
\caption{ 
\small 
  The dependence on the kinematical cut, $z_{min}$, (first column) 
  of the number of data points with $z > z_{min}$ analyzed in the 
  fits (second column) and the resulting $\chi^2/N_{\mathrm{dat}}$ 
  in the {\tt SHKS22} fit set I at NLO. 
}
\label{tab:scan}
\end{table*}

\begin{figure*}[t!]
\resizebox{0.480\textwidth}{!}{\includegraphics[angle=-90]{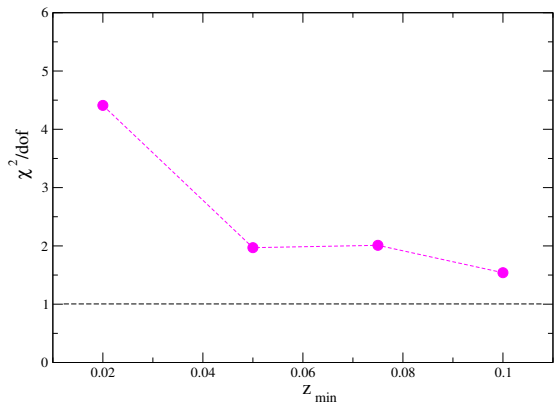}} 
\begin{center}
\caption{ 
\small 
  Dependence of the fit quality, $\chi^2/N_{\mathrm{dat}}$, 
  at NLO on the minimum cut values of $z$ for the SIA datasets 
  used in the {\tt SHKS22} analysis for set I. }
\label{fig:scan}
\end{center}
\end{figure*}


We select data for which a description in the framework of 
perturbative QCD can be expected to work well and exclude 
the range of very small values of $z$. Higher-order corrections 
are known to contain potentially large logarithms proportional 
to $\ln z$ and $\ln (1-z)$. Since the available data are 
limited to the range $z<0.5$ we do not have to consider 
a cut for the large $z$ region. However, we have to carefully 
choose a lower kinematical cut for $z$. In order to do so, we 
study the sensitivity of $\chi ^2/N_{\mathrm{dat}}$ to the 
variations of $z_{min}$ at NLO accuracy. We scan $z_{min}$ 
over the region $0.02 \leq z \leq 0.1$. The summary of our 
selection of $z_{min}$ is presented in Table.~\ref{tab:scan}. 
The first and second columns of this table show our choice of 
$z$ and the remaining number of data points, respectively. 
The total number of data without imposing the cut is $36$ 
points and all of these data points are used in the analysis 
with $z_{min}=0.02$ cut. In the third column, we report  
$\chi^2/N_{\mathrm{dat}}$ determined from the analyses with 
different numbers of data and minimum $z$ cuts.
 
In Fig.~\ref{fig:scan}, the dependence of $\chi ^2/N_{\mathrm{dat}}$ 
on the minimum cut value of $z$ is presented. One can conclude 
from this figure that the optimal $\chi^2/N_{\mathrm{dat}}$ value 
at NLO is obtained by a fit to data with $z_{min}=0.1$. We find 
that there is no further improvement on $\chi ^2/N_{\mathrm{dat}}$ 
by increasing the number of data points and reducing $z_{min}$ 
to values below $0.1$. We choose the cut $z_{min} = 0.1$ for 
all experiments, independent of the center-of-mass energy. 
The number of data points remaining after applying this cut 
is $27$. 

We note that the low-$z$ cut, introduced to remove the region 
where potentially large logarithms from higher-order corrections 
can spoil the reliability of the theoretical predictions,  
allows us as well to omit corrections due to the non-zero 
hadron mass. Such corrections are expected to be relevant for 
small center-of-mass energies and in the small-$z$ region.

%
\section{FF Sets and Fit Quality}
\label{sec:results}
%

\begin{figure*}[htb]
\vspace{0.50cm}
\resizebox{0.480\textwidth}{!}{\includegraphics{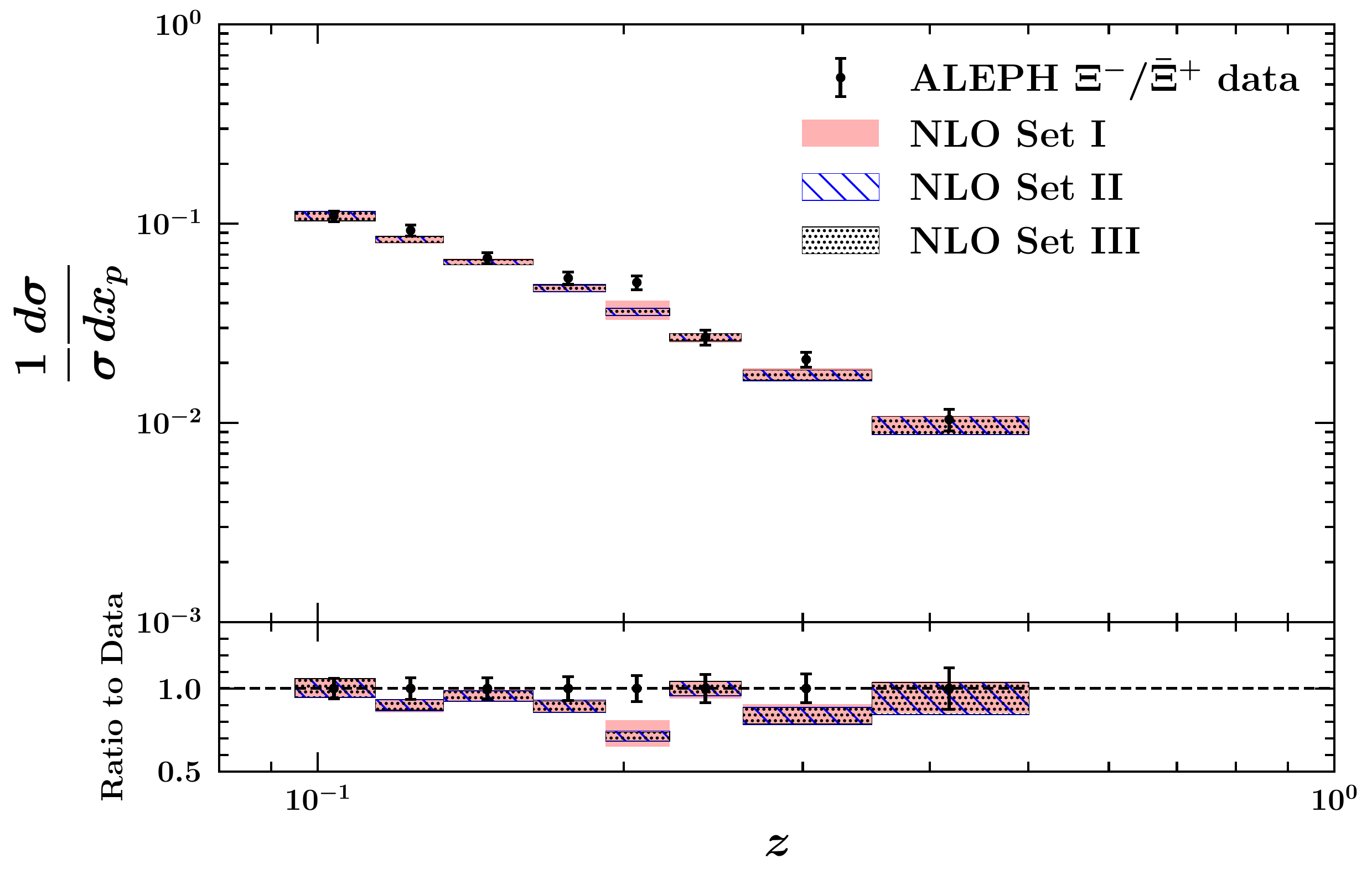}} 	
\resizebox{0.480\textwidth}{!}{\includegraphics{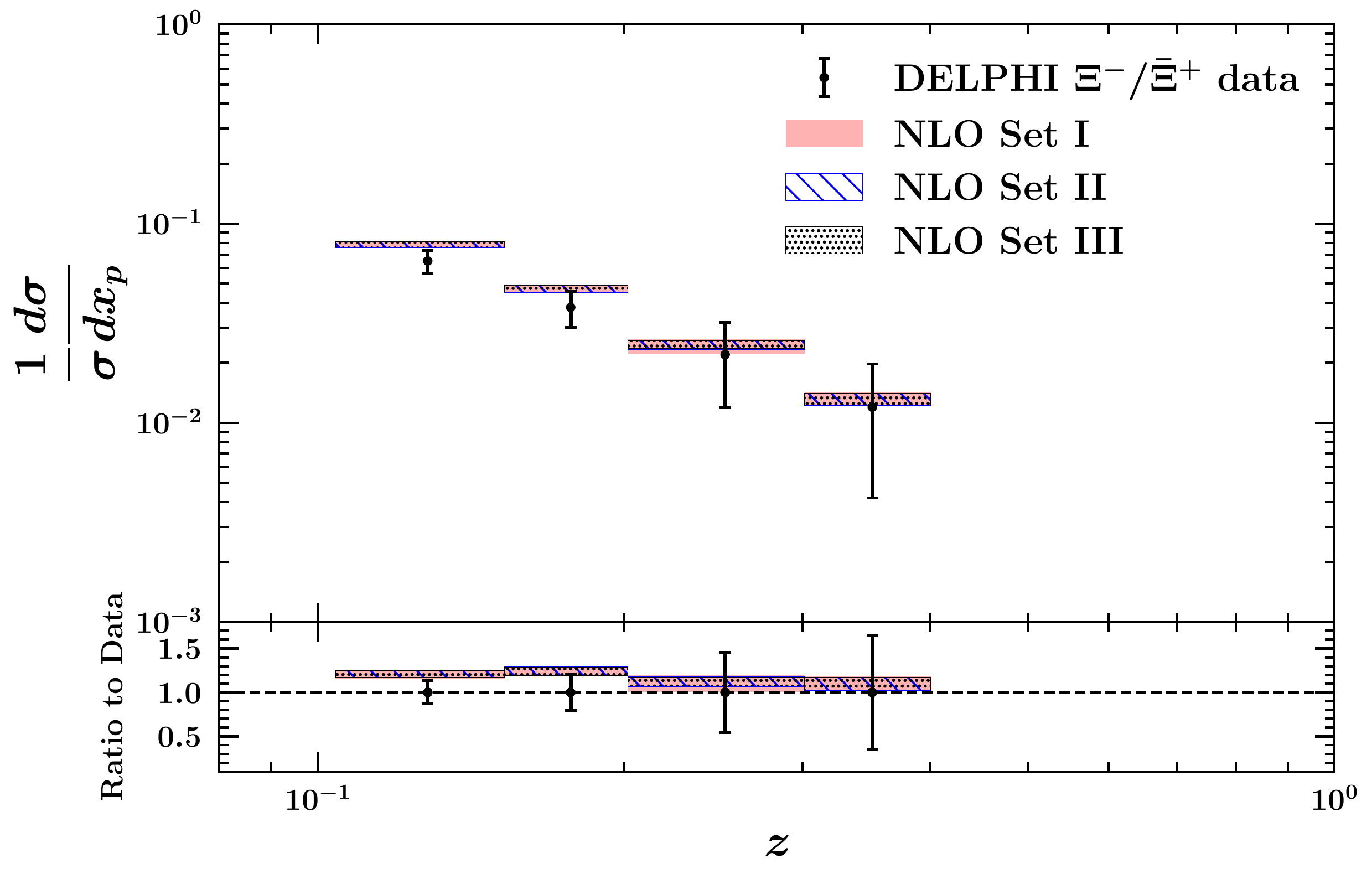}}  	
\resizebox{0.480\textwidth}{!}{\includegraphics{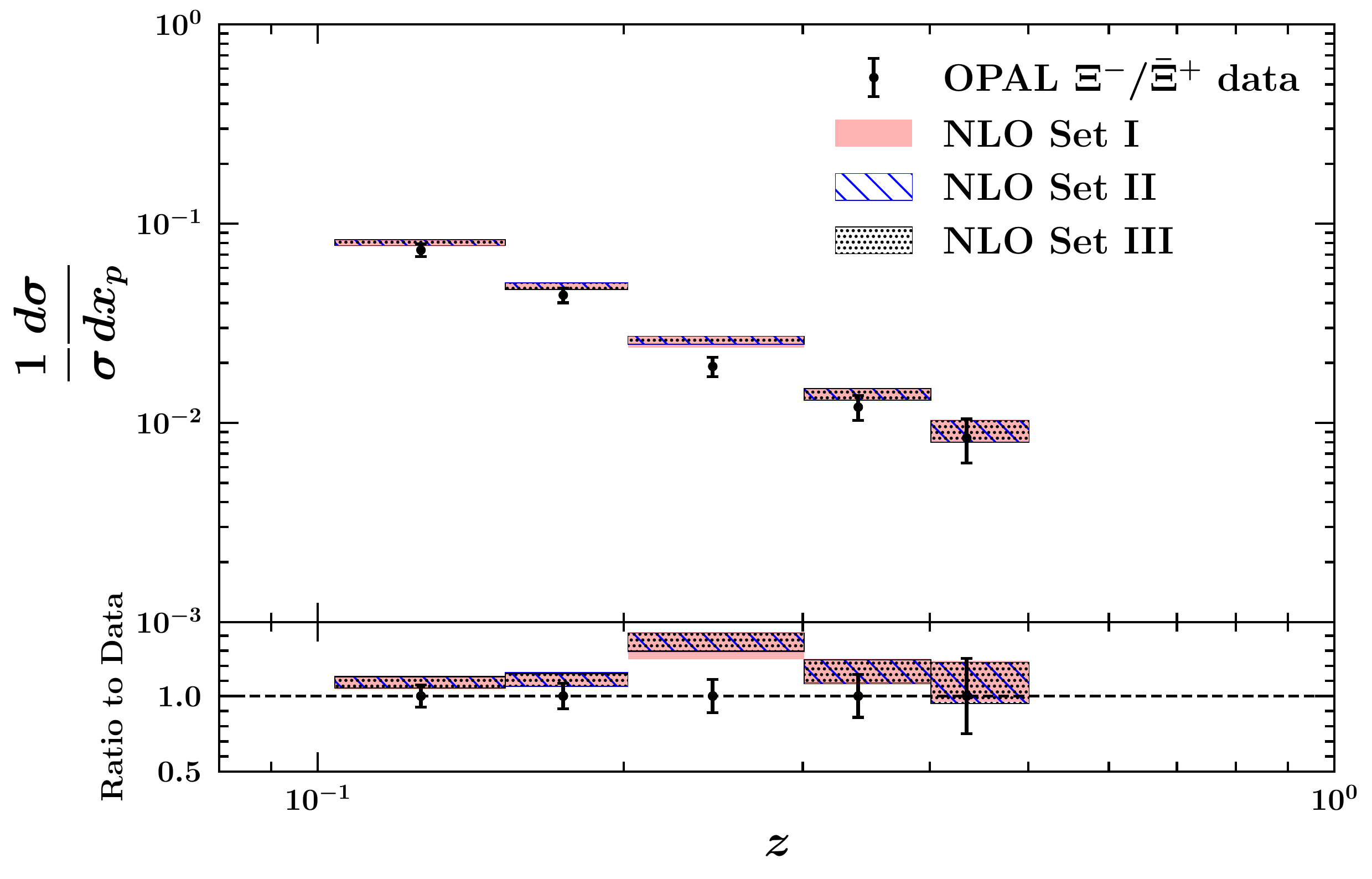}} 	 	
\resizebox{0.480\textwidth}{!}{\includegraphics{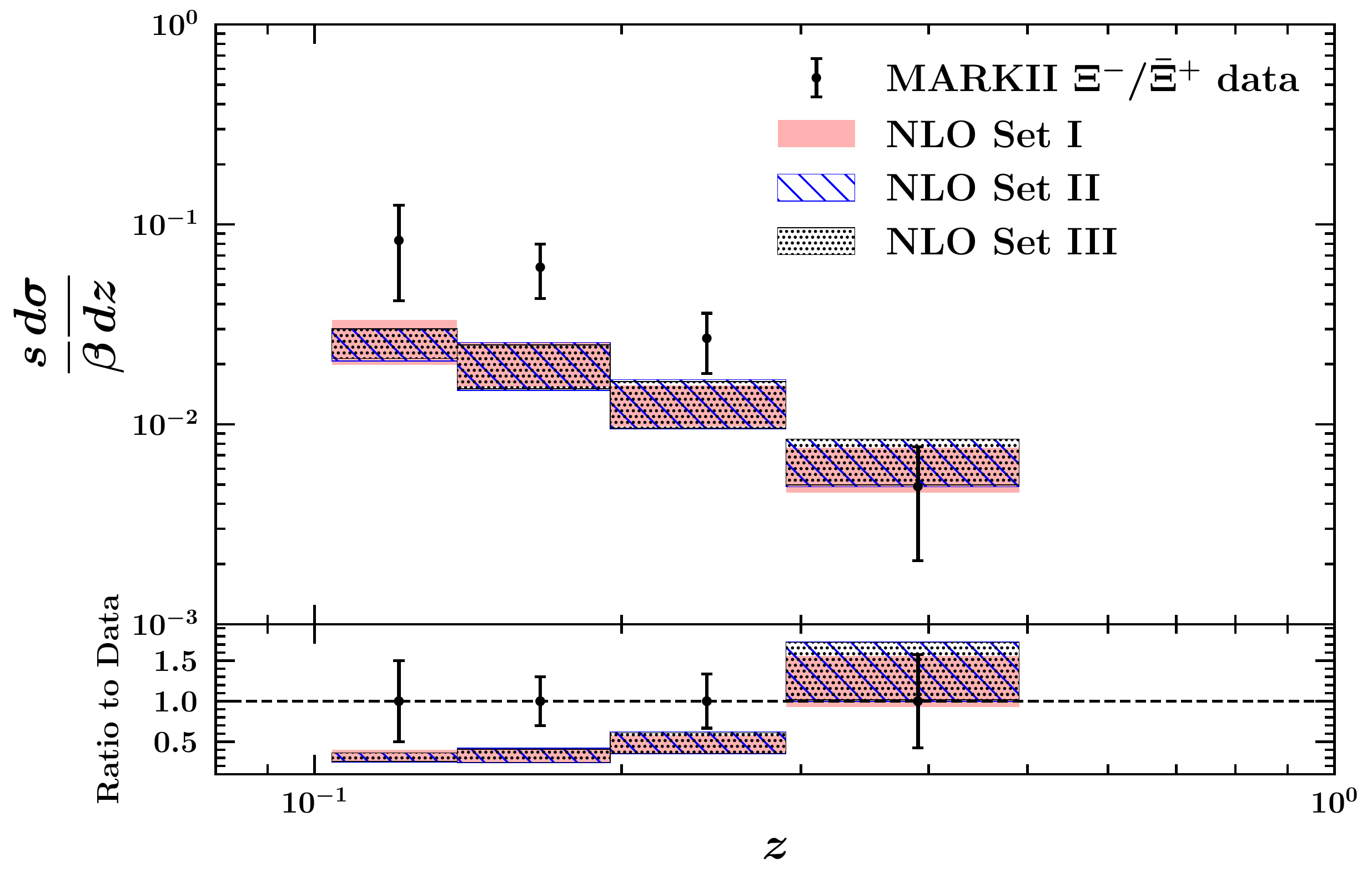}}
\resizebox{0.480\textwidth}{!}{\includegraphics{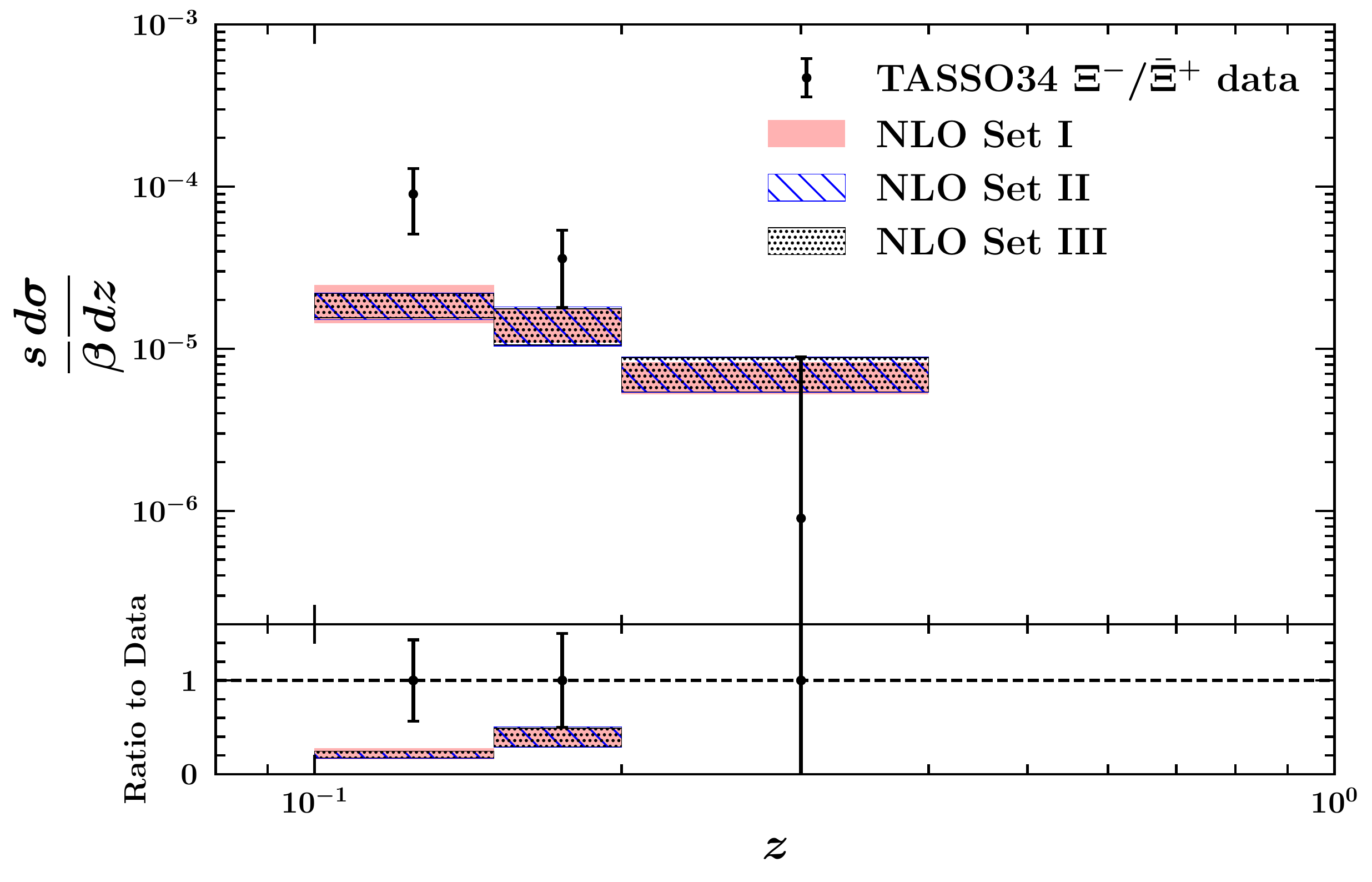}} 	
\resizebox{0.480\textwidth}{!}{\includegraphics{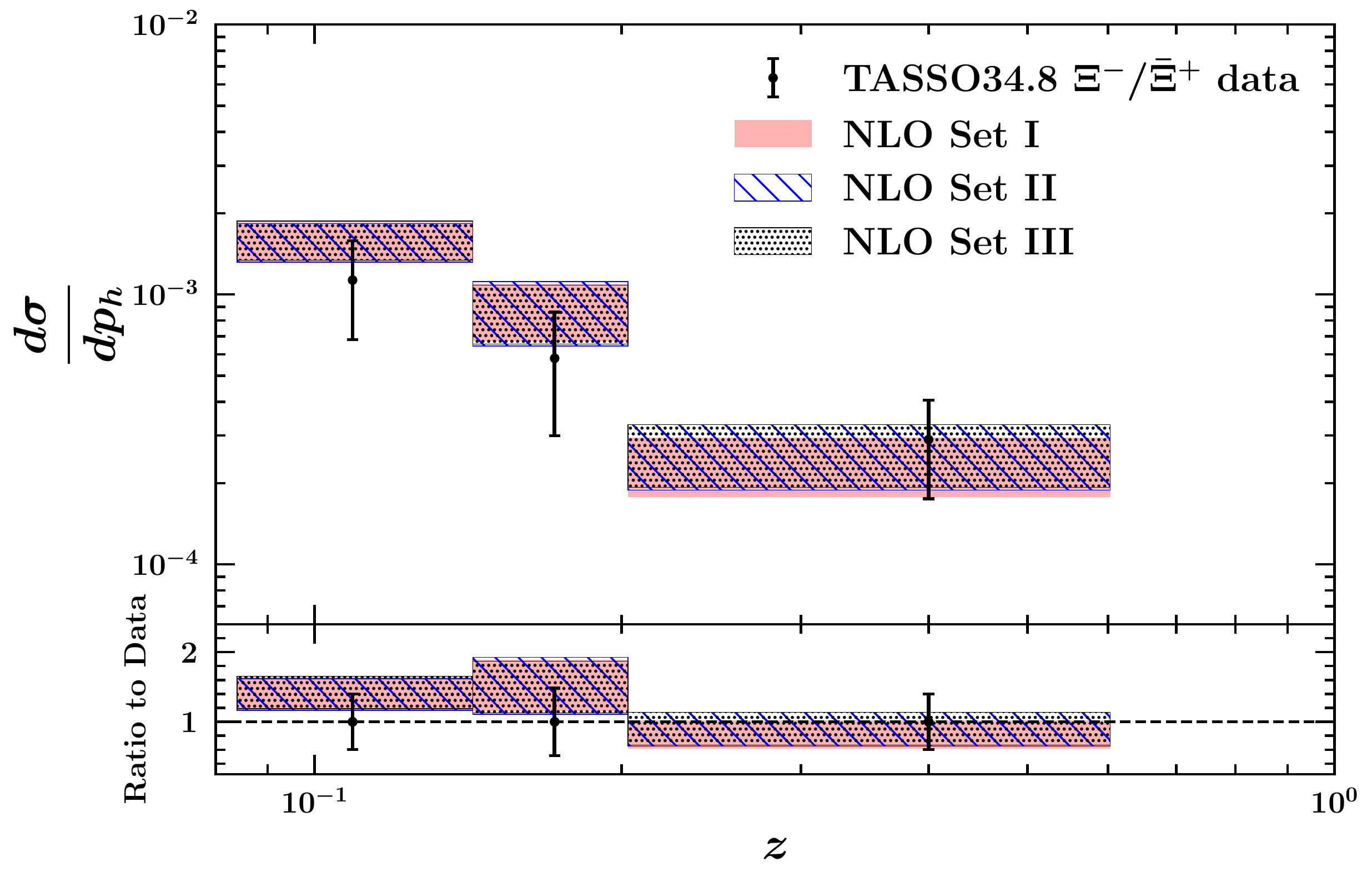}} 	 	
\begin{center}
\caption{ \small 
NLO theory predictions and data/theory ratios for the normalized 
SIA cross section data. See text for details. 
}
\label{fig:data/theory}
\end{center}
\end{figure*}

In this section we describe the main results of the {\tt SHKS22} 
analysis. We present a comparison of the $\Xi ^-/\bar{\Xi}^+$ 
experimental data used in our analysis with the corresponding 
theoretical predictions calculated using our extracted 
{\tt SHKS22} NLO FFs and we also present a comparison of the 
fits at NLO and NNLO. 

A detailed comparison of theoretical predictions obtained using 
the NLO FFs of set I with the SIA experimental data included 
in the fit is shown in Fig.~\ref{fig:data/theory}. The data 
are shown as a function of $z$, limited to the range 
$0.1 \leq z < 0.5$ which corresponds to the cuts imposed in 
our selection of data to be fitted. We present both absolute 
values in upper panels and the ratios of theory prediction 
over data in lower panels. The results are shown for the 
ALEPH~\cite{ALEPH:1996oqp}, DELPHI~\cite{DELPHI:1995dso}, 
OPAL~\cite{OPAL:1996gsw},  MARKII~\cite{Klein:1986ws} and 
the TASSO Collaborations~\cite{TASSO:1983qye,TASSO:1988qlu}. 
Overall, one can see a very good agreement between our theory 
predictions and the measurements for most of the data points. 
We observe some deviations for the MARKII and TASSO 
data at $\sqrt{s} = 34$~GeV. These data points contribute 
substantially to $\chi ^2$. For the case of ALEPH data, the 
agreement between theory predictions and data looks good; 
however, since the experimental errors are typically small 
their contribution to the total $\chi^2$ is also large.

In Table~\ref{tab:datasets}, introduced already above, $\chi^2$ 
values per data point for each individual dataset are shown. 
The total $\chi^2$ divided by the number of data points for 
each set is shown in the last row. All fits are acceptable. One 
can observe a slight increase of $\chi^2 / N_{\mathrm{dat}}$ 
when we go from set I to sets II and III. We have included 
the results for sets II and III in Fig.~\ref{fig:data/theory}. 
A closer inspection of these figures shows that the differences 
between the three sets is mainly due to two data points, one 
of the ALEPH and one of the OPAL data. These points appear as 
outliers and their contribution to the change of $\chi^2$ 
values between the three sets is particularly large. 

\begin{figure*}[t!]
\resizebox{0.480\textwidth}{!}{\includegraphics{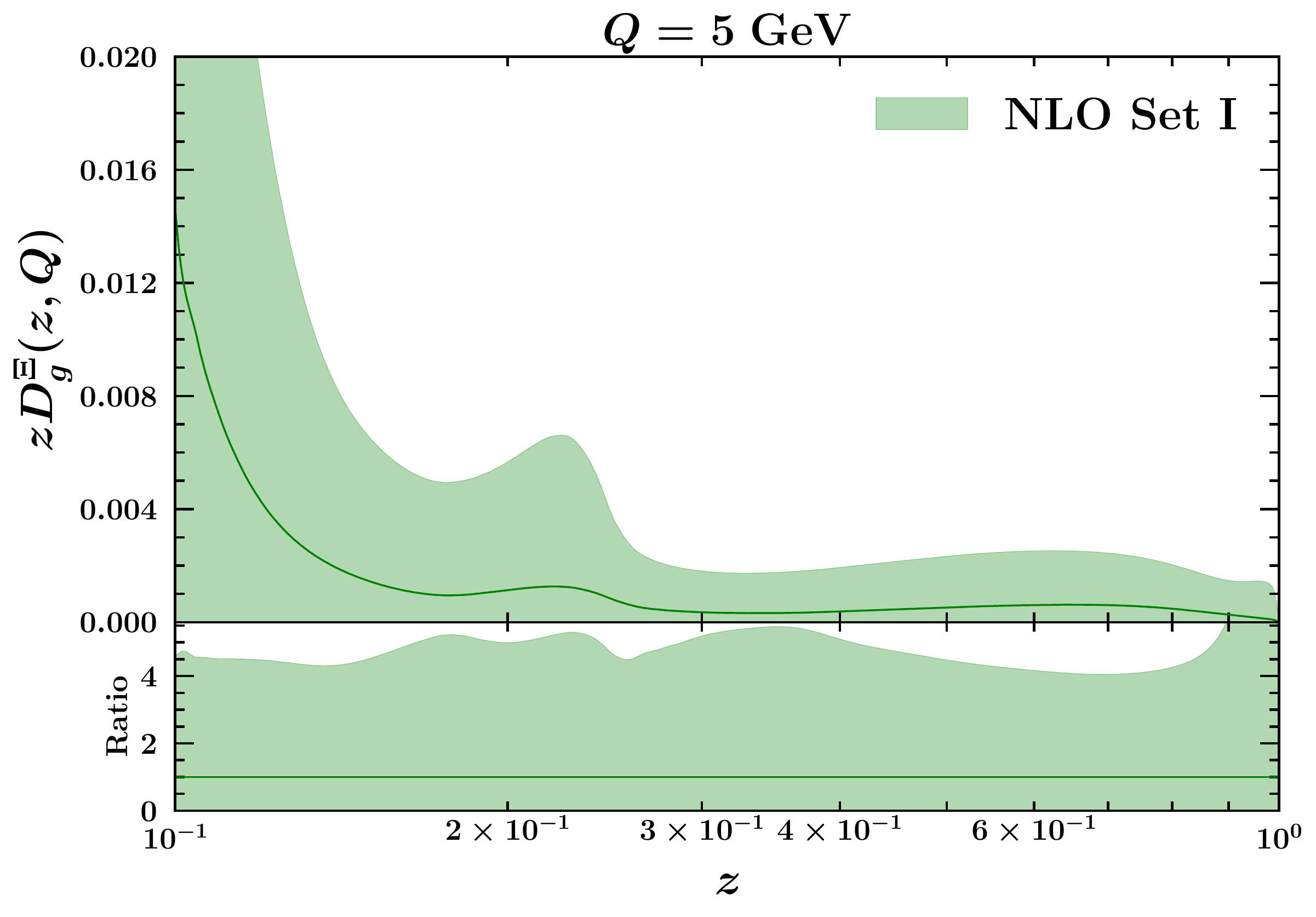}} 	
\resizebox{0.480\textwidth}{!}{\includegraphics{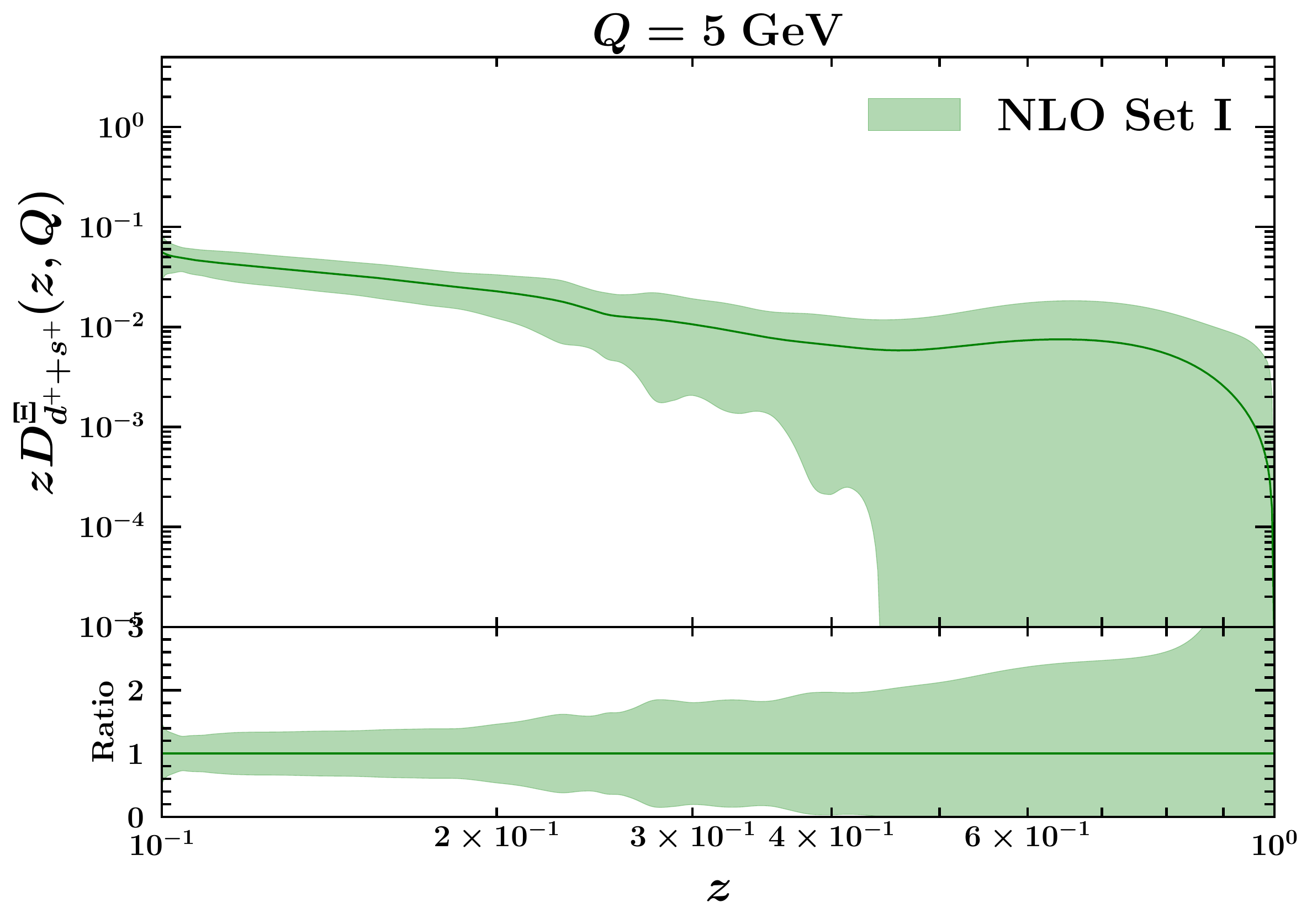}}  	
\resizebox{0.480\textwidth}{!}{\includegraphics{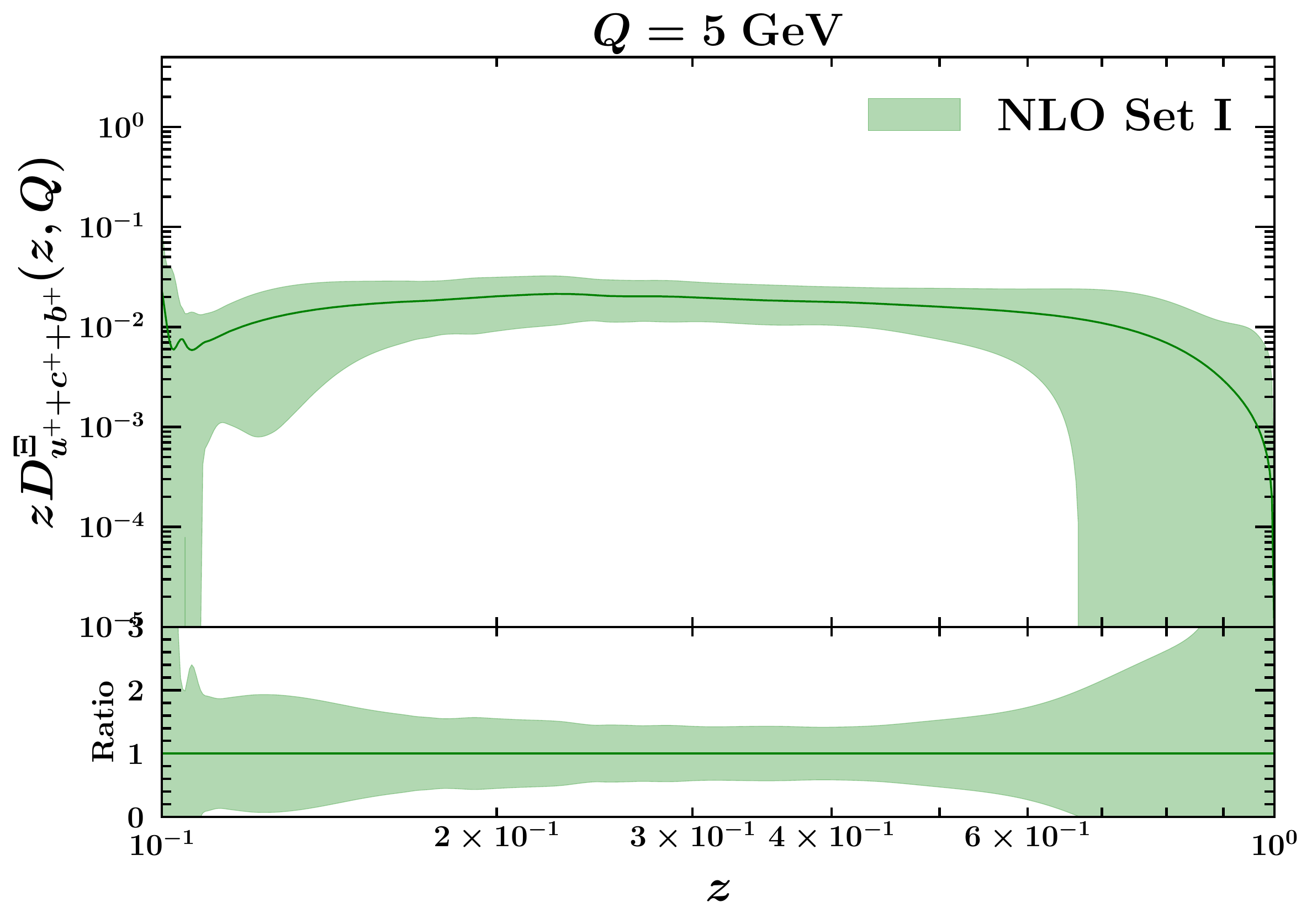}} 
\begin{center}
\caption{ \small 
The flavor combinations of set I FFs at NLO for $\Xi ^- + 
\bar{\Xi}^+$ production. Full lines describe central values, 
shaded areas represent corresponding uncertainty bands at the 
initial scale $Q_0=5$~GeV. The lower panels in each figure 
show the error bands normalized to the corresponding central 
values. 
}
\label{fig:FF3fl}
\end{center}
\end{figure*}

The NLO FFs of $\Xi ^- + \bar{\Xi}^+$ for set I are shown in 
Fig.~\ref{fig:FF3fl} together with the corresponding uncertainty 
bands for the two flavor combinations $d^+ + s^+$ and 
$u^+ + c^+ + b^+$ and the gluon at the initial scale $Q_0=5$~GeV.

\begin{figure*}[htb]
\vspace{0.50cm}
\resizebox{0.480\textwidth}{!}{\includegraphics{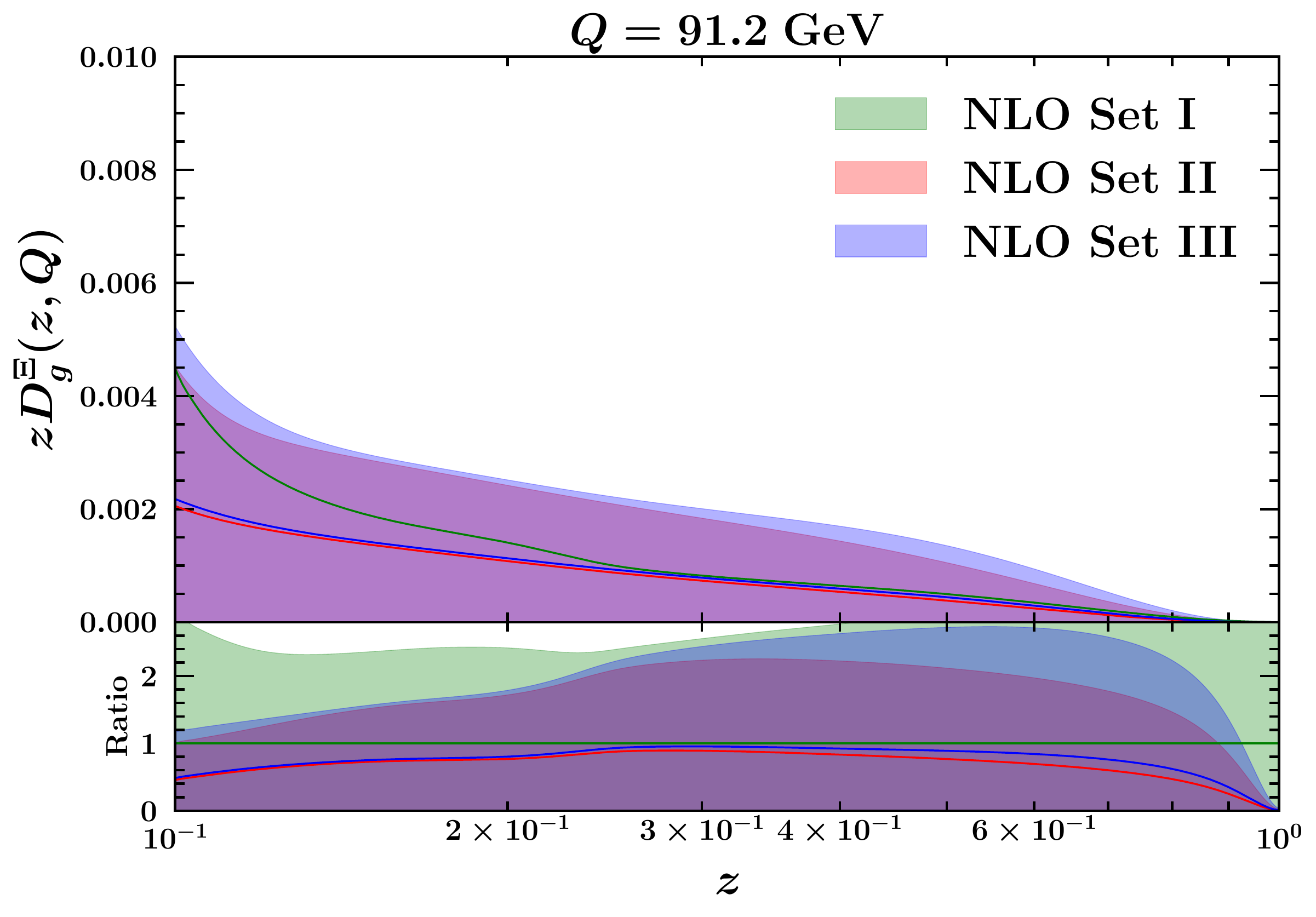}} 	
\resizebox{0.480\textwidth}{!}{\includegraphics{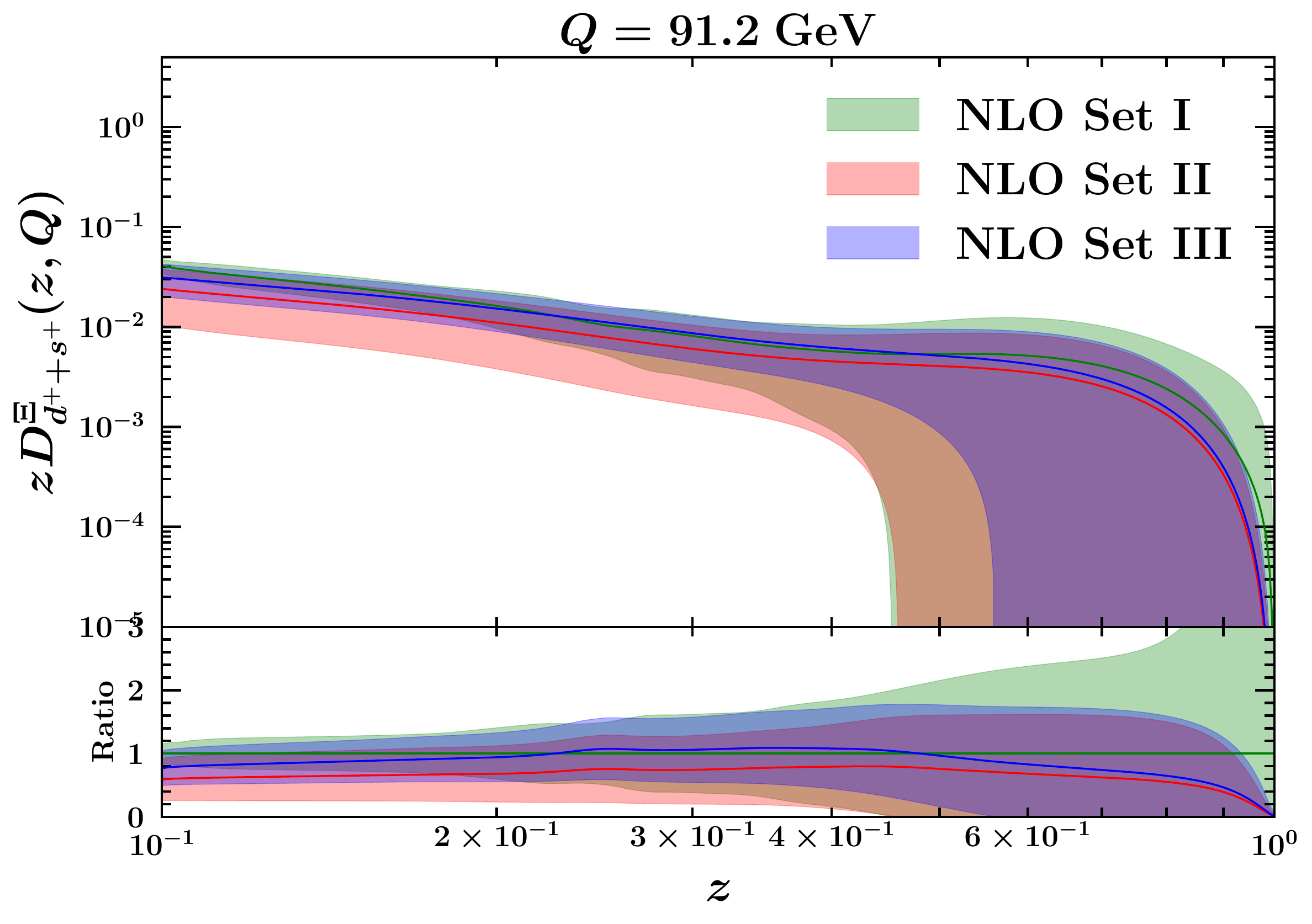}}  	
\resizebox{0.480\textwidth}{!}{\includegraphics{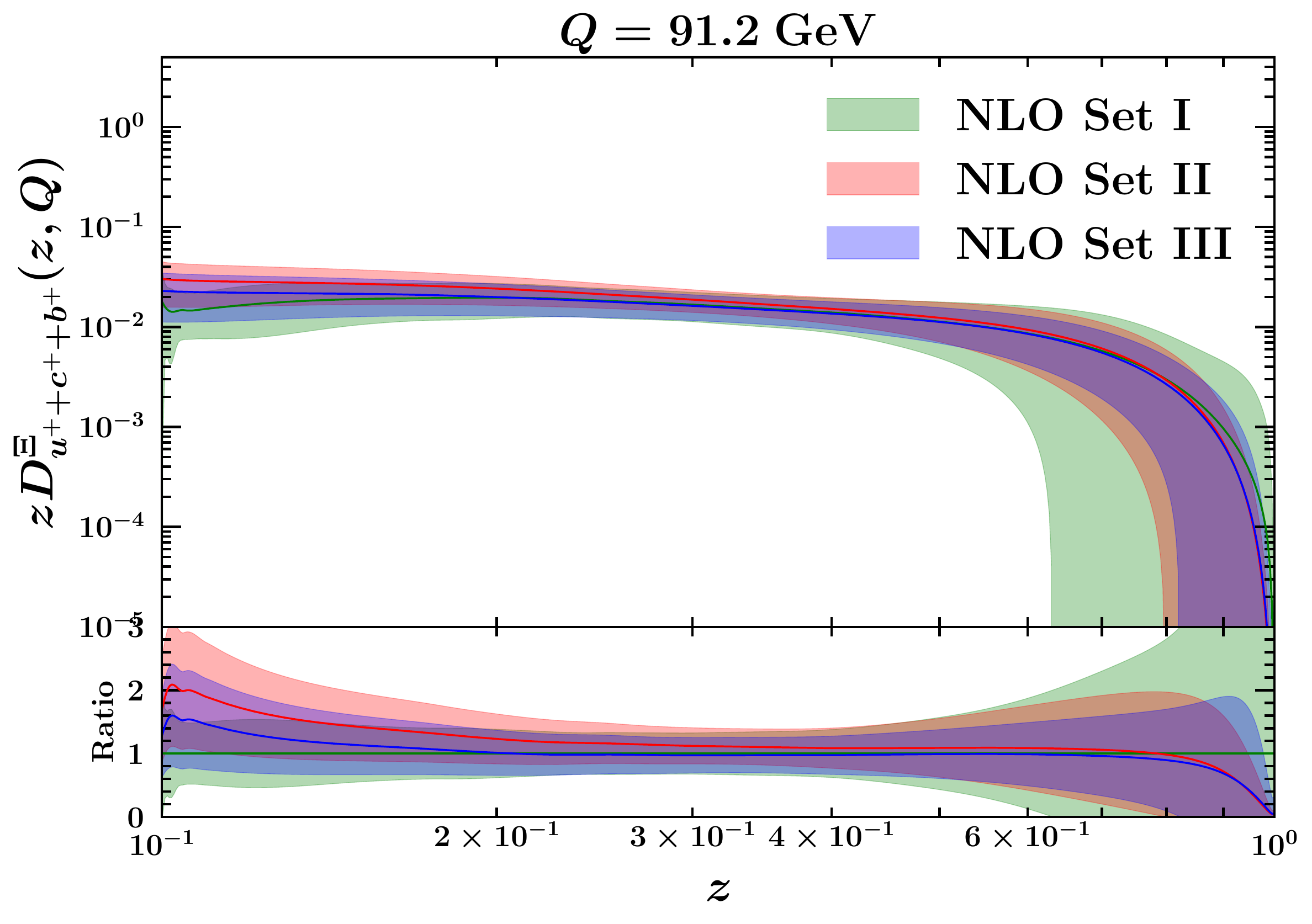}} 
\begin{center}
\caption{ \small 
Comparison of the FFs of set I, II and III at the scale 
$Q = M_Z = 91.2$~GeV. We display the NLO results for 
$\Xi ^- + \bar{\Xi}^+$ FFs, $zD^{\Xi ^- + \bar{\Xi}^+}_i (z)$ 
($i = g$, $d^+ + s^+$, $u^+ + c^+ + b^+$).
}
\label{fig:FF356}
\end{center}
\end{figure*}

A comparison of the three sets of FF fits is shown in 
Fig.~\ref{fig:FF356}. Here we display the NLO results for 
the $g$, $d^+ + s^+$, and $u^+ + c^+ + b^+$ FFs at the 
scale $Q = M_Z$. One can see a remarkable agreement between 
these different sets and the differences between the central 
values are almost invisible in the plots. The uncertainty bands, 
however, for the $d^+ + s^+$ and $u^+ + c^+ + b^+$ combinations 
are wider for set I in the range of $z > 0.4$, while these FF 
combinations are better constrained for set I in the small 
$z$-range. In fact, the available data points cover only the 
range $z < 0.4$ and FF values above this range are the result 
of an extrapolation. We observe that the 
uncertainties of the gluon FF is larger than 
of the quark FFs for all fits. This was to be exptected since 
the SIA data are only loosely sensitive to the gluon FF. 

For sets II and III, i.e.\ when we allow individual flavor 
components of the FFs to be different, we found that the 
central values for the same-charge combinations, i.e.\ 
$d^+$, $s^+$ and $b^+$ on the one hand and $u^+$ and $c^+$ 
on the other hand, are very similar. Therefore we do not 
show corresponding figures. The uncertainty bands, however, 
turn out to be much wider if individual components are fitted, 
as to be expected since the available data do not provide 
enough information for a separate determination of individual 
flavor FFs; only the corresponding sums enter the non-singlet 
part of the structure function $F_2$ in Eq.~(\ref{eq:nonsinglet}). 

The FFs of set I contain only those flavor combinations 
which are needed to describe the available data, which do 
not distinguish light and heavy quark-initiated processes. 
Obviously, this is the appropriate choice to reflect the 
information contained in the data. Therefore we adopt set I 
as our  baseline parametrization and present the comparison of 
NLO and NNLO results only for this set in the following. 

\begin{figure*}[htb]
\vspace{0.50cm}
\resizebox{0.480\textwidth}{!}{\includegraphics{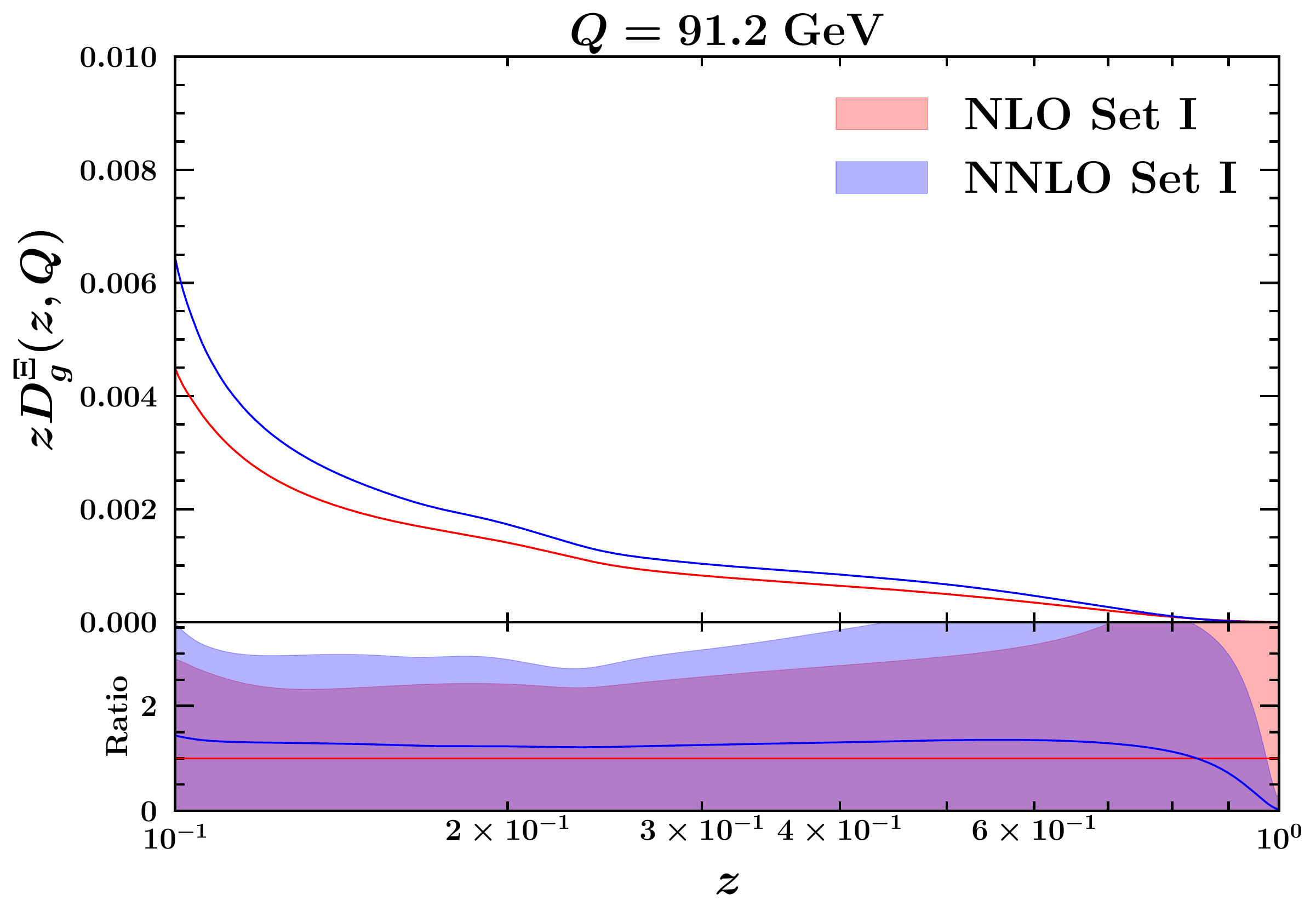}} 	
\resizebox{0.480\textwidth}{!}{\includegraphics{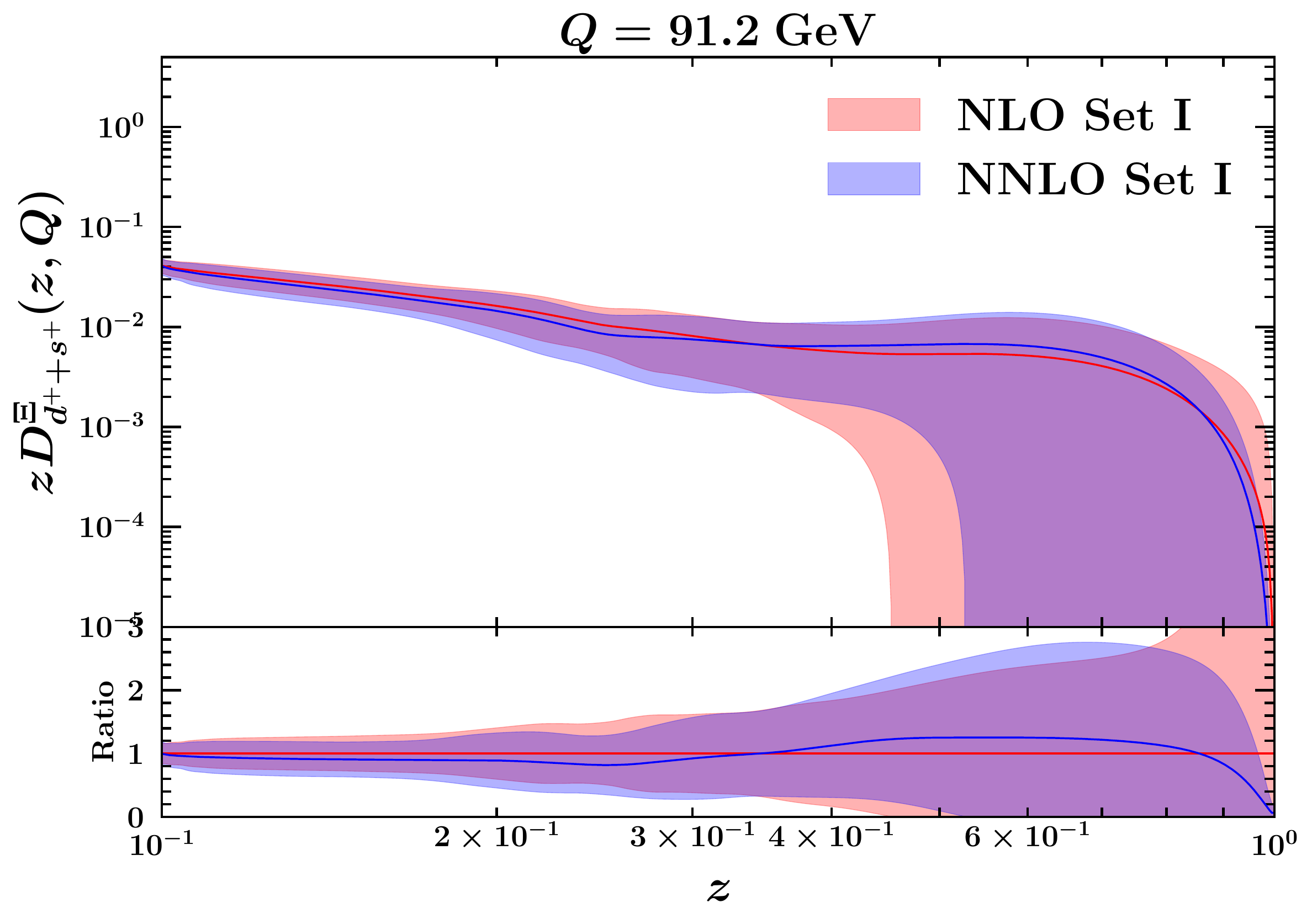}}  	
\resizebox{0.480\textwidth}{!}{\includegraphics{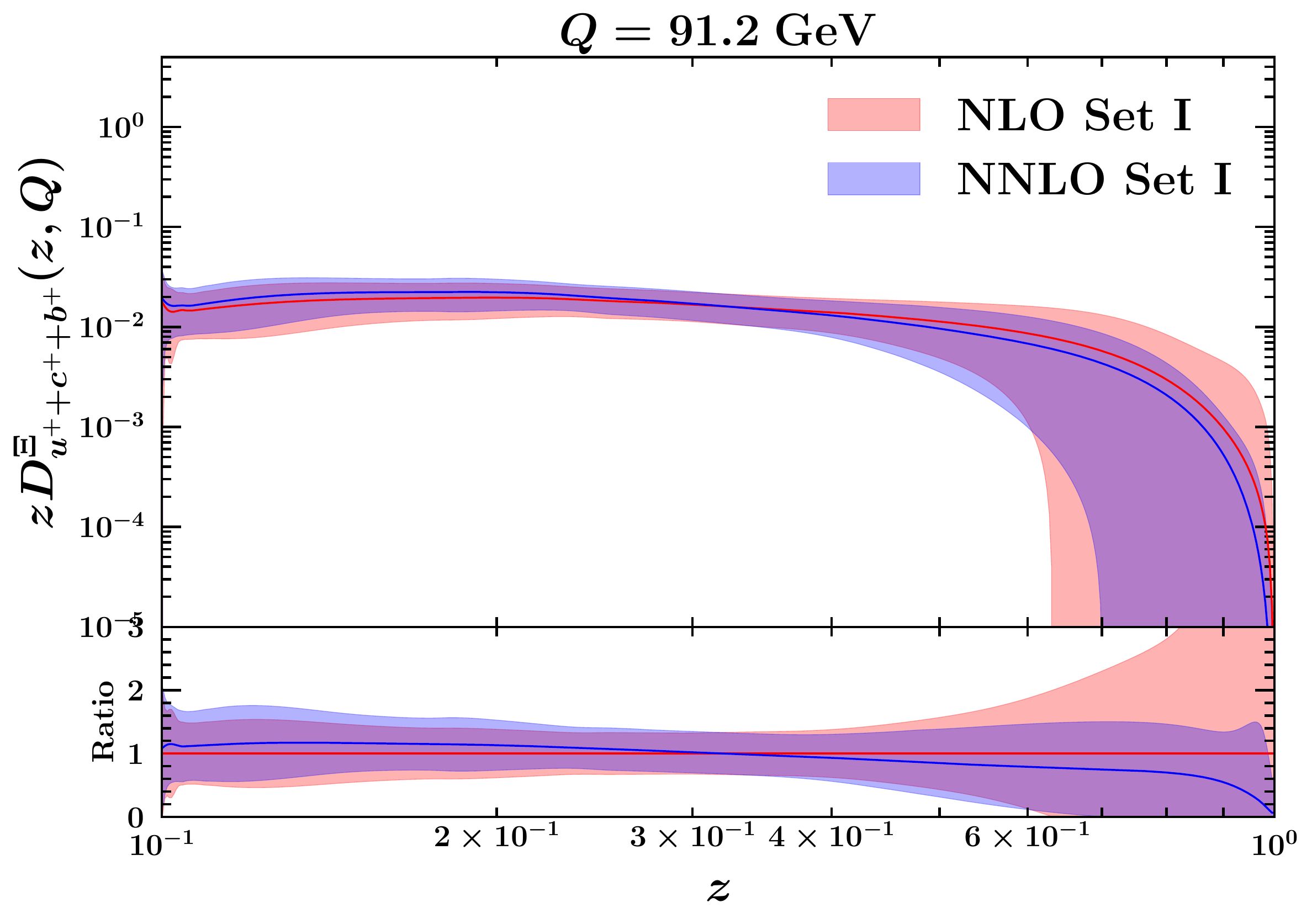}} 
\begin{center}
\caption{ \small 
Comparison of the NLO and NNLO versions of the 
$\Xi ^- + \bar{\Xi}^+$ FFs for the flavor combinations of set I 
at the scale $Q = M_Z$. 
}
\label{fig:FFNLONNLO}
\end{center}
\end{figure*}

A comparison of the NLO and NNLO fits for set I is shown in 
Fig.~\ref{fig:FFNLONNLO}. Central values and uncertainty bands 
of the FFs in Fig.~\ref{fig:FFNLONNLO} differ only little 
between the NLO and NNLO calculations. The error bands overlap 
everywhere. Only in the large-$z$ region, we observe some 
reduction of the uncertainties in the NNLO analysis. The fit 
quality can be judged from the $\chi^2$ values contained in 
Tab.~\ref{tab:datasets}. We observe that the NNLO fit yields 
improved values of $\chi^2/N_{\mathrm{dat}}$: $1.451$ for NNLO 
compared with $1.540$ for the NLO fit, i.e.\ an improvement 
of ${{\sim}} 10\%$. This reduction of $\chi^2$ is shared among 
almost all data points except those of DELPHI and TASSO-34.8 
where we observe a small increase. Overall, the inclusion of 
higher-order perturbative QCD corrections leads to a marginally 
improved set of FFs.  

%
\section{$\Xi^-/\bar{\Xi}^+$ Production at Hadron Colliders} 
\label{sec:LHC-RHIC}
%

Data for $\Xi^-/\bar{\Xi}^+$ production at hadron colliders, 
$p\bar{p}$ scattering at the Tevatron and $pp$ collisions at 
the LHC, are presently available only at very low values of 
the transverse momentum, $p_T$. Perturbative QCD is not expected 
to work reliably in this $p_T$ range and a description in a 
picture assuming independent parton-to-hadron fragmentation is 
possibly missing dominating non-perturbative effects. In order 
to separate such non-perturbative effects and identify possible 
collective phenomena, which could become important for example 
in the hot quark-gluon plasma in heavy-ion collisions, it will 
be important to compare data with predictions in a framework 
using FFs obtained at high energy scales and extrapolating down 
into the $p_T$ range where data already exist. On the other hand, 
future measurements at the LHC experiments at higher $p_T$ can 
eventually contribute to an improved determination of the 
$\Xi^- + \bar{\Xi}^+$ FFs.

Hadron production in $pp$ and $p\bar{p}$ collisions are 
sensitive to relatively large $z$ values while the available 
data from SIA are restricted to $z < 0.5$. To some extent, 
our predictions are therefore based on extrapolating the 
$\Xi^- + \bar{\Xi}^+$ FFs into a yet unconstrained kinematic 
range. We have checked, however, that the hadron collider 
cross sections described below do have some overlap with the 
SIA data: about 20\,\% of the $pp$ and $p\bar{p}$ cross sections 
originate from the region with $z < 0.5$.

Here we show results of the calculation of cross sections for 
collider experiments. 
 
\begin{figure*}[htb]
\vspace{0.50cm}
\resizebox{0.480\textwidth}{!}{\includegraphics{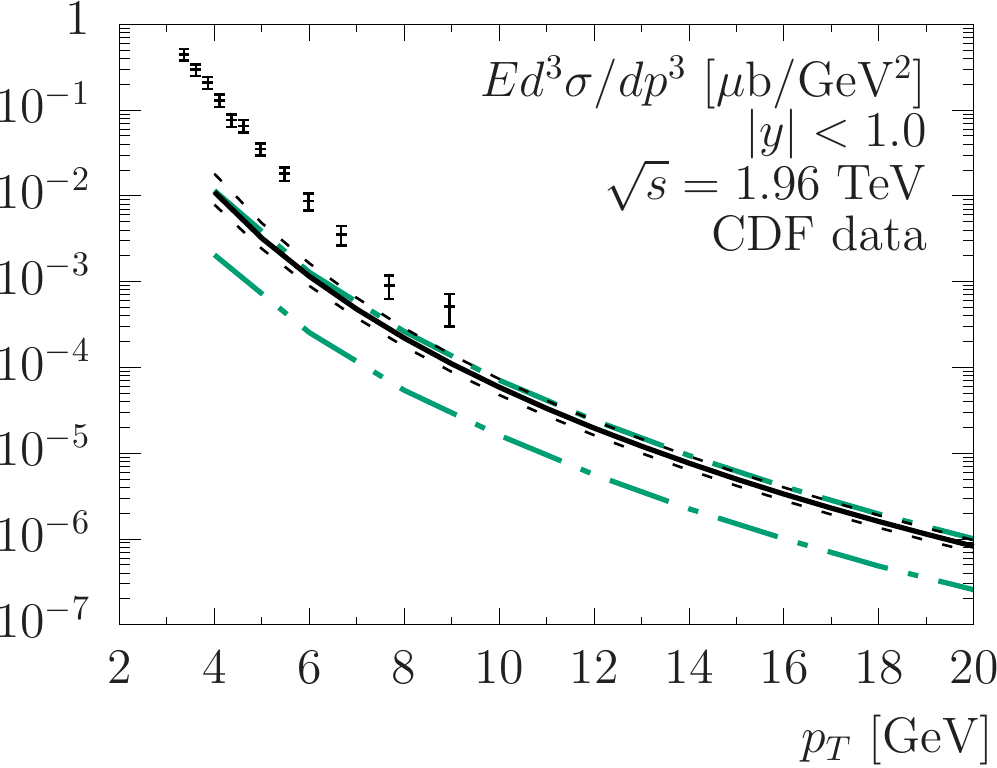}} 
\resizebox{0.480\textwidth}{!}{\includegraphics{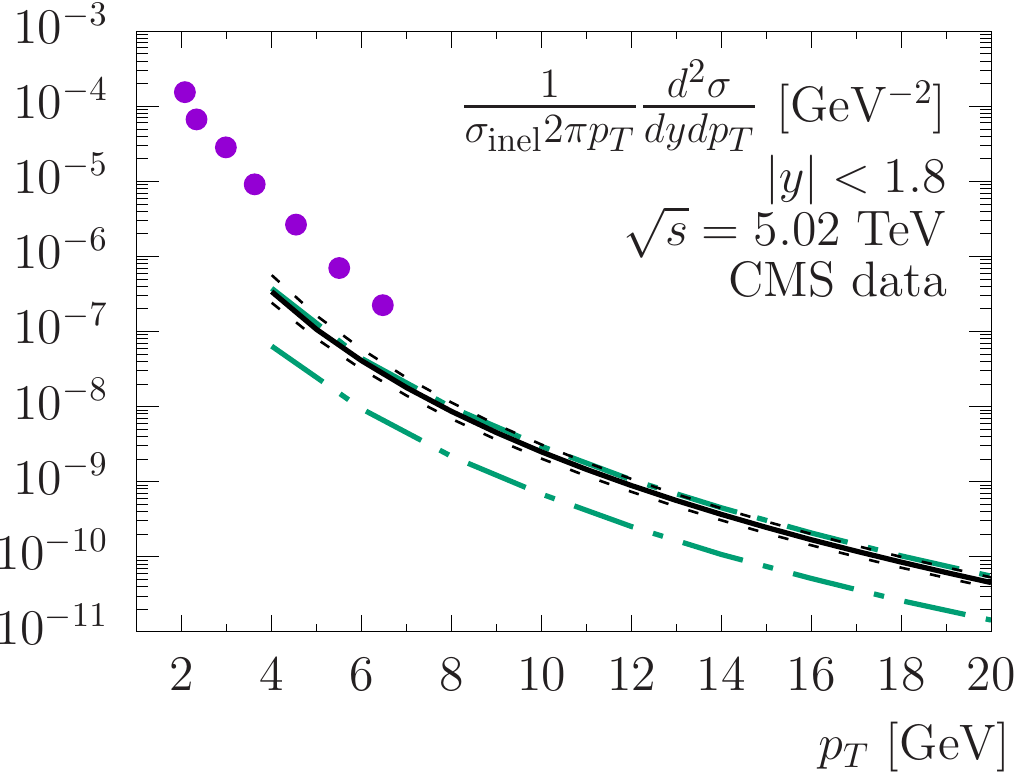}}  
\resizebox{0.480\textwidth}{!}{\includegraphics{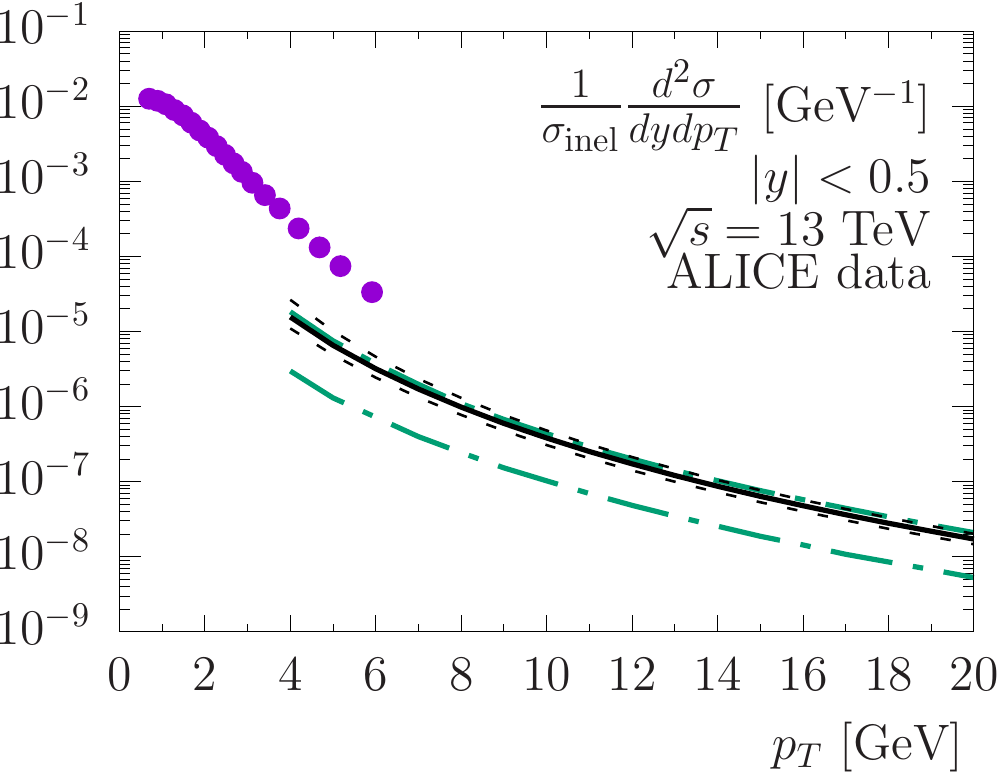}} 
\begin{center}
\caption{\small 
Predictions for $\Xi ^- + \bar{\Xi}^+$ production at hadron 
colliders: CDF at the Tevatron \cite{CDF:2011dvx}, 
CMS \cite{CMS:2019isl} and ALICE \cite{ALICE:2020jsh} 
at the LHC. In the case of CMS and ALICE, data points have 
been read off from the corresponding published figures and 
error bars are not shown since they are too small to be reliably 
extracted from the figures. 
Parton distributions from \cite{Hou:2019qau,Hou:2019efy}, 
set CT18nlo, were used and theory errors are shown as described in 
the text. 
}
\label{fig:sigmacollider}
\end{center}
\end{figure*}

The calculation is done with code that was originally developped 
for the production of heavy quarks ({\tt gmvfns-3.4}, described 
in Ref.~\cite{Kniehl:2005mk}), but adapted for massless quarks 
for the present purpose. It is based on the results of the 
calculation of Ref.~\cite{Aversa:1988vb} which includes 
next-to-leading order corrections and takes into account all 
sub-processes with incoming and outgoing quarks and gluons. 
In order to connect our results with the 
existing data, we compare with measurements in $p\bar{p}$ 
collisions from CDF~\cite{CDF:2011dvx} at $\sqrt{s} = 1.96$~TeV 
in the rapdity range $|y| \leq 1$, as well as in $pp$ collisions 
from CMS~\cite{CMS:2019isl} at $\sqrt{s} = 5.02$~TeV, $|y| \leq 
1.8$ and from ALICE \cite{ALICE:2020jsh} at $\sqrt{s} = 13$~TeV, 
$|y| \leq 0.5$. Results are shown in Fig.~\ref{fig:sigmacollider}, 
compared with the existing data in the $p_T$ range up to about 
6 -- 8~GeV and predictions extending up to $p_T = 20$~GeV. The 
default predictions, using the CT18nlo set of parton distributions 
from \cite{Hou:2019qau,Hou:2019efy} and the central member of the 
set I $\Xi^- + \bar{\Xi}^+$ FFs turn out to be considerably 
smaller than the data, as expected. 

There are two important sources for uncertainties of the 
theoretical predictions: variations of the renormalization and 
the factorization scales, and uncertaintites due to the FF parametrization. Renormalization and factorization scales have 
been set equal to $p_T$ in our default predictions. The 
short-dashed lines in Fig.~\ref{fig:sigmacollider} show the 
error band due to variations of the renormalization scale by 
factors of 2 up and down. The uncertainty from variations of 
the factorization scales is always smaller than that due to 
variations of the renormalization scale. The uncertainty from 
the FF parametrization is, however, dominating. This is shown 
by the long-dash-dotted (green) lines in Fig.~\ref{fig:sigmacollider}. 
The corresponding error band is obtained by scanning over 300 
replicas of the $\Xi^- + \bar{\Xi}^+$ FFs. The predicted cross 
sections follow an asymmetric distribution with a mean value 
much smaller than the result from the central FF. The error 
band shown in Fig.~\ref{fig:sigmacollider} covers the 
$1\sigma$-range, obtained by excluding the lowest and highest 
$16\%$ of predicted cross section values \cite{Butterworth:2015oua}. 
Obvsiously, with FFs from our present {\tt SHKS22} analysis, 
there are very large uncertainties. In turn, this means that 
future measurements at the LHC can be expected to provide 
valuable information to further constrain the FF parametrization 
for $\Xi^- + \bar{\Xi}^+$ hadron production.  
This will be particularly important for the gluon FF for which 
our fits with presently available data resulted in large 
uncertainties. In contrast to SIA data, the measurements at 
hadron colliders are dominated by the gluon-to-$\Xi$ FF: 
about 70\,\% of the cross section is due to this FF component. 
Therefore one can expect considerably reduced uncertainties 
of the gluon FF from future LHC data.

%
\section{Summary and Conclusions} 
\label{sec:conclusion}
%

In this paper we have introduced a new set of unpolarized FFs 
for $\Xi ^-/\bar{\Xi}^+$ production, called {\tt SHKS22}. 
This analysis presents the first precise determination of 
$\Xi ^-/\bar{\Xi}^+$ fragmentation functions which include 
higher-order perturbative corrections at NNLO accuracy. 
The current analysis is based on a comprehensive dataset from 
single inclusive electron-positron annihilation experiments. 
We found that including data points in the region of $z \geq 0.1$ 
allows us to obtain stable fit results. 

We have explored three different FF sets and parametrized them 
in terms of Neural Networks. This approach is expected to lead 
to a substantial reduction of theoretical bias in the choice 
of a parametrization. The parameters describing the Neural 
Networks are fitted to the SIA data. We have used the recent 
publicly available package {\tt MontBlanc} to set up the 
parametrization of FFs, perform their evolution and calculate 
the SIA production cross-sections. In addition, a Monte Carlo 
sampling method is used to propagate the experimental 
uncertainties into the fitted FFs and to calculate the 
uncertainties for FFs and corresponding obervables.

All three sets of FFs obtained in this work describe the SIA 
experimental data reasonably well. Our baseline set I contains 
a minimal choice of parameters describing the sum of favored 
quark flavors $d^+ + s^+$ and assumes symmetry between the 
disfavored quark flavors $u^+$, $c^+$ and $b^+$. The available 
data do not provide enough information to obtain precise FFs 
for each quark flavor separately. 

Our new results provide an important ingredient needed for 
the calculation of theoretical predictions for the hadron 
production in proton-proton collisions. We have used our 
FFs to calculate the inclusive $\Xi ^- + \bar{\Xi}^+$ 
baryon production in proton-proton collisions, extending 
the presently covered transverse momentum range of the 
CDF, CMS and ALICE measurements. Corresponding future 
measurements can then be used to extract $\Xi ^- + \bar{\Xi}^+$ 
FFs in combination with SIA data.  

The main result of this work, i.e.\ the $\Xi ^- + \bar{\Xi}^+$ 
FFs, are available as grid files formatted for the {\tt LHAPDF} 
library from the authors upon request.

%
\begin{acknowledgments}
%

The authors are thankful to Valerio Bertone for many helpful 
discussions and comments. M.S., H.H. and H.K. thank the School 
of Particles and Accelerators, Institute for Research in 
Fundamental Sciences (IPM) for financial support provided for 
this research. M. S. is thankful
to the Iran Science Elites Federation for the financial support.
H.K. is thankful to the Department of Physics at the University of 
Udine and International Centre for Theoretical Physics (ICTP) for the financial support provided for this research.

\end{acknowledgments}

\clearpage


\end{document}